\documentclass[twocolumn,preprintnumbers,amsmath,amssymb]{revtex4}  

\usepackage{graphicx}
\usepackage{dcolumn}
\usepackage{bm}

\begin{document}

\title{Topological Phase Transitions and a Two-Dimensional Weyl Superconductor in a Half-Metal/Superconductor Heterostructure}

\author{Lei Hao$^{1,2}$ and C. S. Ting$^{1}$}
 \address{$^1$ Texas Center for Superconductivity and Department of Physics,
University of Houston, Houston, Texas 77204, USA  \\$^2$ Department of Physics, Southeast University, Nanjing 210096, China}

\date{\today}

\begin{abstract}
We find a series of topological phase transitions in a half-metal/superconductor heterostructure, by tuning the direction of the magnetization of the half-metal film. These include transitions between a topological superconducting phase with a bulk gap and another phase without a bulk gap but has a ubiquitous local gap. At the same time, the edge states change from counter-propagating Majorana edge modes to unidirectional Majorana edge modes. In addition, we find transitions between the second phase and a nodal phase which turns out to be a two-dimensional Weyl superconductor with Fermi line edge states. We identify the topological invariants relevant to each phase and the symmetry that protects the Weyl superconductivity phase.
\end{abstract}


\maketitle

\section{\label{sec:Introduction}Introduction}

The discovery of topological insulator has infused great enthusiasm in finding new materials with novel topological
properties.\cite{hasan10,qi11} One particularly exciting subject is how to realize Majorana fermions, which is not only of fundamental interest but also has potential application in fault-tolerant
topological quantum computation.\cite{nayak08} Among many schemes proposed up-to-date\cite{read00,ivanov01,fu08,sau10,lutchyn10,alicea10,oreg10,qi10,mourik12,choy11,perge13,perge14}, the heterostructure consisting of a semiconductor clamped
between a magnetic insulator and an $s$-wave superconductor appears to be the most promising one\cite{sau10}. The spin-splitting of the Fermi surface of the semiconductor by the Rashba spin-orbit
coupling (RSOC) and the existence of an out-of-plane Zeeman field are  essential ingredients of the scheme. Similar mechanism has also been proposed in cold atom systems.\cite{zhang08,sato09} An
alternative system, a heterostructure consisting of a half-metal (HM) and an $s$-wave superconductor ($s$SC) containing the same essential physical ingredients has also been studied.\cite{lee09,chung11}

In existing hybrid solid-state systems proposed to host the topological superconducting phase and the Majorana fermions, the magnetization (exchange field) is required to be perpendicular to the spin
of the charge carriers fixed by the SOC.\cite{fu08,sau10,lutchyn10,alicea10,chung11} In the HM/$s$SC heterostructure, however, there is an intrinsic degree of freedom in directing the magnetization of the HM
thin film, as shown schematically in Fig.1. This is achieved either by cutting the thin film along different high symmetry directions of the bulk parent material, or through a magnetic field when the
magnetic anisotropy of the parent HM is small. For an ideal HM without SOC, there exists $SU(2)$ symmetry with respect to the simultaneous rotation of the magnetization and the electron spin. The
above tunability generates no physical difference as the exchange field or the magnetization changes its directions. However, the formation of heterostructure with the substrate and the
superconductor enforces the inversion asymmetry to the HM thin film along the normal direction of the film. The ensuing RSOC breaks the above $SU(2)$ symmetry.\cite{rashba} In the presence of the
RSOC, the physics associated with a general orientation of the magnetization so far has not been examined in the proposed heterostructure. It is thus interesting to know whether there is a single
phase or there are topologically distinct phases for different directions of the magnetization in the HM/$s$SC heterostructure.

\begin{figure}\label{fig1} \centering
\includegraphics[width=7.5cm,height=2.6cm]{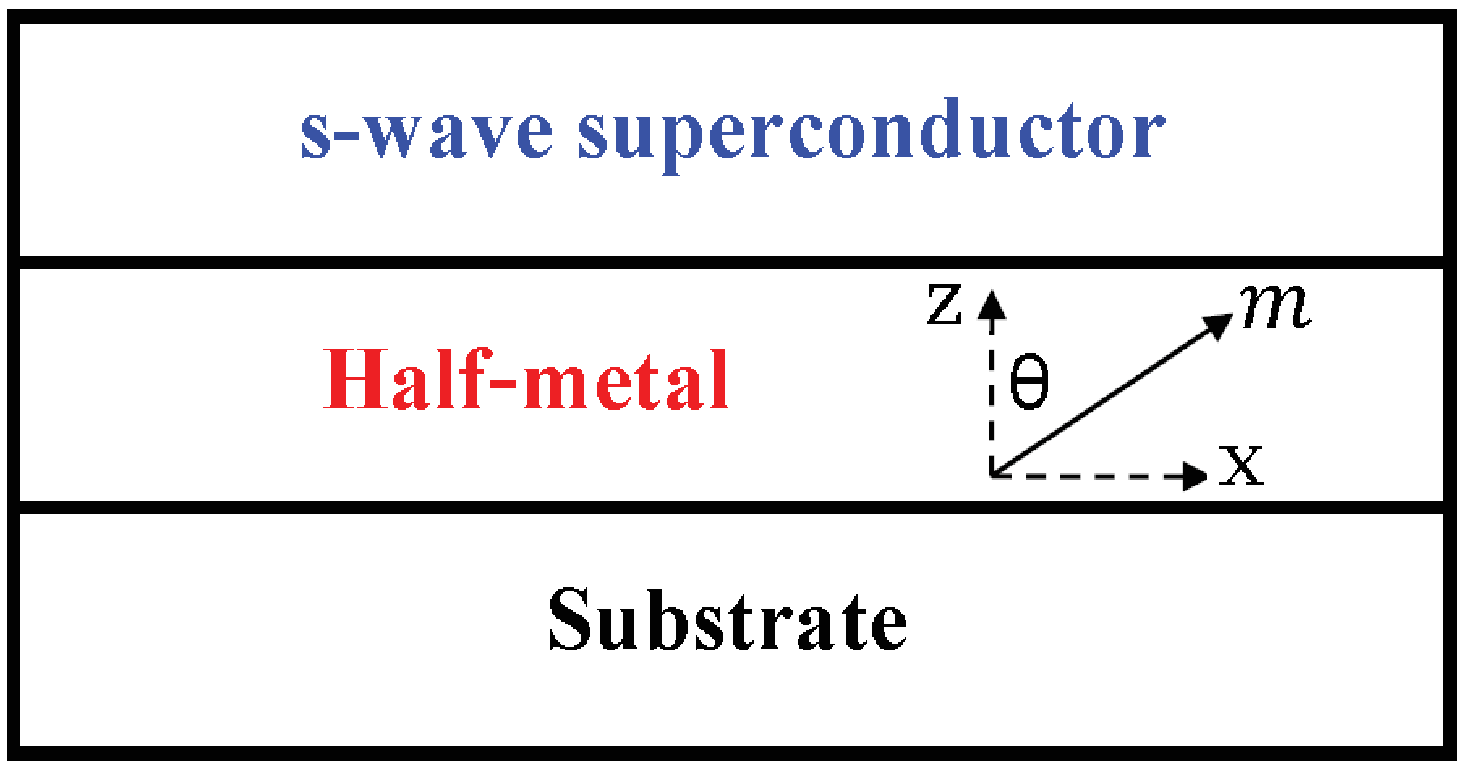}  \\
\caption{Schematic drawing of a heterostructure consisting of a half-metal (HM) thin film sandwiched between an $s$-wave superconductor and an insulating substrate, viewed laterally. $m$ and $\theta$ are the magnitude and
direction of the exchange field (magnetization) in the HM, which is assumed to lie on the $xz$ plane.}
\end{figure}

Inspired by the above observations, we study in this paper the possible topological phases existing in the HM/$s$SC heterostructure. Three phases with distinctive topological numbers and edge states
are found. In particular, we demonstrate that a Weyl superconductivity phase with two Weyl nodes appears in the heterostructure and it is protected by an emergent mirror symmetry of the system when the magnetization lies in the plane of the HM film.

\section{\label{sec:model}Model and symmetries}

To illustrate the principle, we consider the simplest model for the HM/$s$SC heterostructure shown in Fig.1. We describe the HM thin film by a
one-orbital model defined on a square lattice and assume perfect interfaces between the HM and the substrate and the $s$SC. Denoting the basis vector as
$\phi^{\dagger}_{\mathbf{k}}=[d^{\dagger}_{\mathbf{k}\uparrow},d^{\dagger}_{\mathbf{k}\downarrow}]$, the model Hamiltonian for the HM thin film with a RSOC term induced by the formation of the
heterostructure is $\hat{H}_{0}=\sum_{\mathbf{k}}\phi^{\dagger}_{\mathbf{k}}h_{0}(\mathbf{k})\phi_{\mathbf{k}}$, where\cite{sau10,chung11}
\begin{equation}
h_{0}(\mathbf{k})=\epsilon_{\mathbf{k}}\sigma_{0}+m_{x}\sigma_{1}+m_{z}\sigma_{3}+\lambda(\sin k_{x}\sigma_{2}-\sin k_{y}\sigma_{1}).
\end{equation}
$\sigma_{0}$ is the rank-2 unit matrix, $\sigma_{i}$ ($i=1,2,3$) are Pauli matrices in the spin subspace. $\epsilon_{\mathbf{k}}=-2t(\cos k_{x}+\cos k_{y})-\mu$. $t$ is the hopping amplitude, $\mu$ is
the chemical potential, and $\lambda$ is the amplitude of the RSOC. Introducing $m$ ($m>0$) and $\theta$ ($0\le\theta<2\pi$) to denote the magnitude and direction of the magnetization (see Fig.1), we have $m_{x}=m\sin\theta$, $m_{z}=m\cos\theta$. To describe a HM with $\hat{H}_{0}$, we assume that $m$ is of the same order of magnitude as $t$, and is much larger than the externally induced $\lambda$. The chemical potential is to be tuned to make sure that it crosses only with the lower spin-split band of $h_{0}(\mathbf{k})$ (see Fig.4(a) in Appendix B for an illustration of the band structure obtained by solving Eq.(1)), which amounts to $\mu\simeq -4t$. The proximity-induced superconductivity in the HM arising from coupling with an $s$SC is described by
$\hat{H}_{p}=\frac{1}{2}\sum_{\mathbf{k}}\phi^{\dagger}_{\mathbf{k}}\underline{\Delta}(\mathbf{k})\phi^{\dagger}_{-\mathbf{k}}+\text{H.c.}$, where
$\underline{\Delta}(\mathbf{k})=\Delta_{0}(\mathbf{k})i\sigma_{2}$.\cite{fu08,sau10} For the sake of simplicity and without losing generality, we ignore in the following analysis the wave vector dependency of the
pairing amplitude and thus we take $\Delta_{0}(\mathbf{k})=\Delta_{0}$ as a real constant (see also Appendix D).\cite{fu08,sau10} In the Nambu basis,
$\varphi^{\dagger}_{\mathbf{k}}=[\phi^{\dagger}_{\mathbf{k}},\phi^{\text{T}}_{-\mathbf{k}}]$, the full model is written as
$\hat{H}=\frac{1}{2}\sum_{\mathbf{k}}\varphi^{\dagger}_{\mathbf{k}}h(\mathbf{k})\varphi_{\mathbf{k}}$, where
\begin{eqnarray}
h(\mathbf{k})&=&\epsilon_{\mathbf{k}}\tau_{3}\sigma_{0}+m_{x}\tau_{3}\sigma_{1}+m_{z}\tau_{3}\sigma_{3}    \notag \\
&&+\lambda(\sin k_{x}\tau_{3}\sigma_{2}-\sin k_{y}\tau_{0}\sigma_{1})-\Delta_{0}\tau_{2}\sigma_{2}.
\end{eqnarray}
$\tau_{i}$ ($i=1,2,3$) are Pauli matrices in the Nambu space. Diagonalizing Eq.(2) gives the four quasiparticle bands $E_{n}(\mathbf{k})$, with $n$ running from $-2$ to $2$ in an order of increasingly higher energy. Only the two low-energy quasiparticle bands, $E_{\pm1}(\mathbf{k})$, have nontrivial topological properties and will be the main focus of our following discussions (See Fig.4(b) in Appendix B for an illustration of the full band structures obtained by solving Eq.(2)).

Now we list the fundamental symmetries of the model relevant to our following discussions. First of all, a nonzero $m$ breaks the time-reversal symmetry of the model for all $\theta$. However, for all
values of $m$ and $\theta$, the model preserves the particle-hole symmetry, $\Xi^{-1}h(\mathbf{k})\Xi=-h(-\mathbf{k})$. The particle-hole operator is defined as $\Xi=\tau_{1}\sigma_{0}K$, in which $K$ denotes complex conjugation. For $\theta=\pi/2$ and $3\pi/2$, the model has a mirror reflection symmetry, which takes $x\rightarrow -x$. The operator acting on
$h_{0}(\mathbf{k})$ for this mirror reflection symmetry is $M_{x}=i\sigma_{1}$, which gives $M^{-1}_{x}h_{0}(\mathbf{k})M_{x}=h_{0}(-k_{x},k_{y})$. The transformation of the pairing term is
$M^{-1}_{x}\underline{\Delta}(M^{\text{T}}_{x})^{-1}=\underline{\Delta}$. For the full model expressed in the Nambu basis, the mirror symmetry is represented as $\tilde{M}_{x}=i\tau_{3}\sigma_{1}$. Finally, the RSOC breaks the inversion symmetry of the model.

\begin{figure}\label{fig2} \centering
\includegraphics[width=8.5cm,height=12cm]{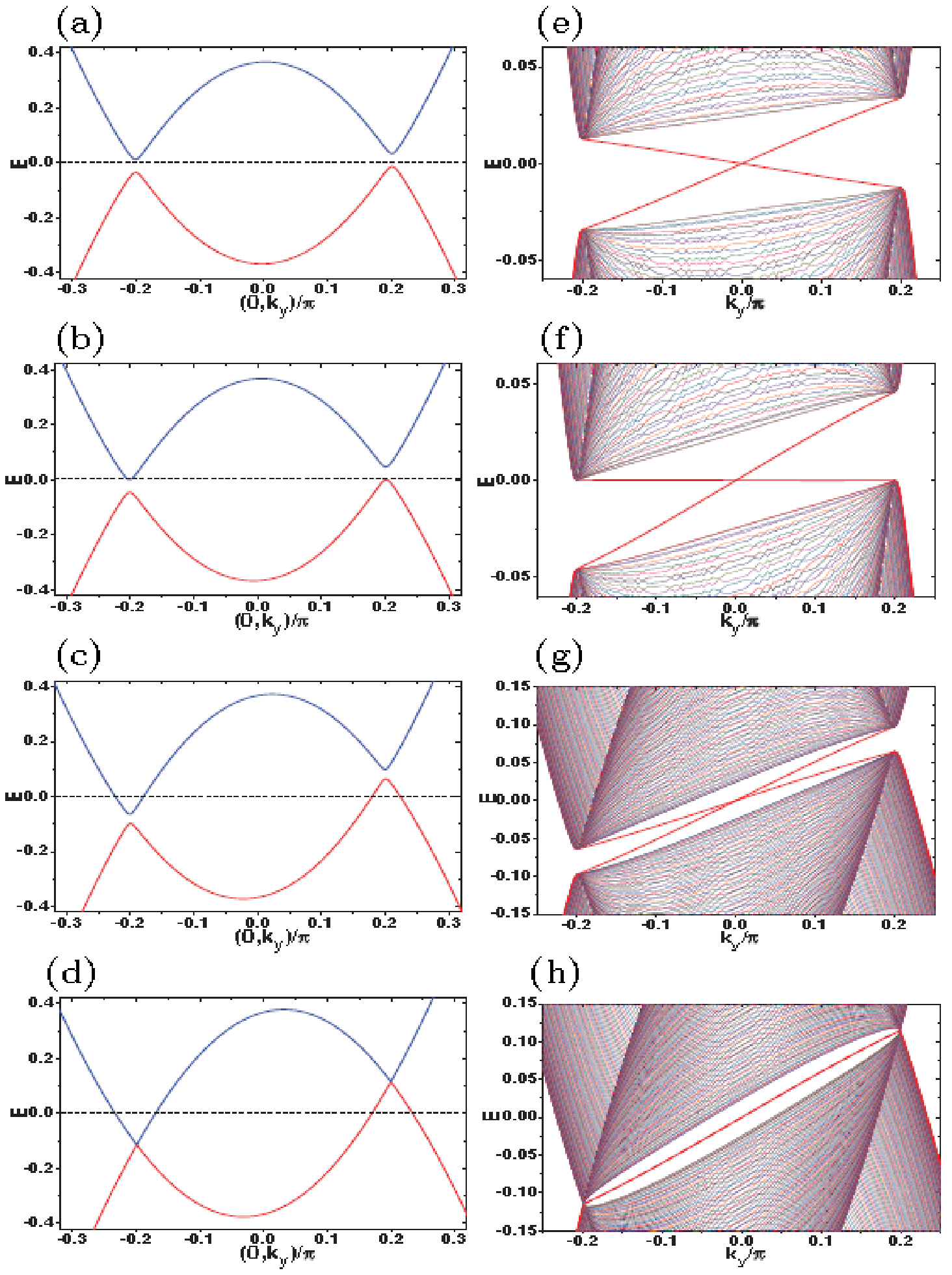}
\caption{Energy spectra for bulk (a, b, c, d) and strips (e, f, g, h) of the system, for a typical set of parameters $m=t>0$, $\lambda=\Delta_{0}=0.2t$, $\mu=-4.6t$. The strips have 500 unit cells along the $x$ direction. $\theta=0.03\pi$ for (a) and (e), $\theta=0.06415\pi$ for (b) and (f), $\theta=0.25\pi$ for (c) and (g), $\theta=0.5\pi$ for (d) and (h). The energies are in unit of $t$.}
\end{figure}

\section{\label{sec:TPT}Topological phase transitions}

To probe possible phases in the HM/$s$SC heterostructure with different orientations of the magnetization, we have calculated the
energy spectra for both the bulk material and a strip of 500 unit cells with two edges running along the $y$ direction. The calculations are based on Eq.(2). Qualitative changes in the bulk and edge state spectra are obtained when the value of $\theta$ is sweeping from 0 to $2\pi$. Focusing on the first quadrant of the cycle ($\theta\in[0,\pi/2]$), Fig. 2 shows results for a typical set of parameters and several particular values of $\theta$. Only the two low-energy quasiparticle bands ($E_{-1}(\mathbf{k})$ and $E_{1}(\mathbf{k})$) are shown. When $\theta\simeq0$, the bulk is fully gapped (Fig.2(a)) and the edge state consists of two counter-propagating modes (Fig.2(e)) which are
known as chiral Majorana fermions.\cite{sau10,alicea10} Increase $\theta$ to a parameter-dependent critical value, $\theta_{c}$, the two low-energy quasiparticle bands are still separated by a local gap in the whole two-dimensional (2D) Brillouin zone (BZ) but the global gap disappears (Fig.2(b)). Correspondingly, one of the two chiral edge modes becomes flat (Fig.2(f)). This critical point $\theta_{c}$ is determined approximately in the limit of $|\lambda|/m\ll1$ and $|\Delta_{0}|/m\ll1$ by
\begin{equation}
|\tan{\theta_{c}}|=\frac{|\Delta_{0}|}{m}.
\end{equation}
See Appendix B for more details of the derivation of Eq.(3). Increasing $\theta$ further, the energy overlap between the two low-energy quasiparticle bands increases (Fig.2(c)), and the two chiral edge modes become unidirectional (Fig.2(g)). Then at $\theta=\pi/2$, not only the bulk gap is absent, the local gap also closes at two nodes along the $(0,k_{y})$ direction (Fig.2(d)), and the two edge modes become degenerate (Fig.2(h)). Since the two low-energy quasiparticle bands are nondegenerate, the appearance of the two nodes and the conelike dispersion close to them (see Appendix B for more details) indicates the showing up of a 2D Weyl superconductivity phase.\cite{wan11,burkov11,murakami07,yang14} The presence of only two Weyl nodes at different energies are consistent with the fact that both time-reversal symmetry and inversion symmetry are broken.\cite{zyuzin12,hosur13,chang15} The phase changes in the other ranges of $\theta$ are qualitatively similar. More spectral properties of the edge states which are relevant to the experimental detection of various phases can be found in Appendix C.

In what follows, we identify the underlying bulk topological invariants relevant to the phase transitions found above. The properties of the Weyl superconductivity phase will be analyzed later (see also Appendix B). In an earlier work by Ghosh \emph{et al}, a Pfaffian $Z_{2}$ invariant for the 2D semiconductor heterostructure was introduced from the particle-hole symmetry.\cite{ghosh10} Since
our system is also 2D and particle-hole symmetric, the same Pfaffian $Z_{2}$ invariant can be defined, which turns out to be
\begin{equation}
P=\text{sgn}[\Delta^{2}_{0}+\epsilon^{2}_{\mathbf{k}=(0,0)}-m^{2}],
\end{equation}
where $\epsilon_{\mathbf{k}=(0,0)}=-4t-\mu$, the function $\text{sgn}(x)$ gives the sign of a real number $x$. See Appendix A for more details on the derivation of $P$. The phase is nontrivial (trivial) if $P=-1$ ($P=1$). Clearly, the above Pfaffian invariant depends only on the magnitude $m$ of the exchange field and is blind to the angle $\theta$. According to Eq.(4), we would have a single topological phase for all $\theta$ once $m>\sqrt{\Delta^{2}_{0}+\epsilon^{2}_{\mathbf{k}=(0,0)}}$.\cite{ghosh10} This is different from what we predicted for the global-gapless phases in Figs. 2(g) and 2(h). Therefore we have to find some finer criteria, if any, to discriminate the different phases in Fig.2.

Because time-reversal symmetry is broken in the present 2D system, a natural topological invariant to consider is the Chern number. One way of calculating the Chern number is through the TKNN
(Thouless-Kohmoto-Nightingale-den Nijs) formula \cite{tknn}
\begin{equation}
C=\frac{1}{2\pi}\iint_{BZ} d^{2}\mathbf{k}\cdot\boldsymbol{\nabla}_{\mathbf{k}}\times\mathbf{A}(\mathbf{k}),
\end{equation}
where the Berry connection is defined as $\mathbf{A}(\mathbf{k})=i\sum_{E_{n}(\mathbf{k})<0}<u_{n}(\mathbf{k})|\boldsymbol{\nabla}_{\mathbf{k}}u_{n}(\mathbf{k})>$, with $|u_{n}(\mathbf{k})>$ the
eigenvector of the $n$-th quasiparticle band. We call $C$ obtained from Eq.(5) as the TKNN number, which gives the Hall conductance of the model (in unit of $e^{2}/h$). Since a local gap exists
between all consecutive pairs of the four quasiparticle bands for all $\theta$ except $\pi/2$ and $3\pi/2$, we are motivated to define another set of Chern numbers related to the respective
quasiparticle bands. Define the Berry connection for the $n$-th quasiparticle band as $\mathbf{A}_{n}(\mathbf{k})=i<u_{n}(\mathbf{k})|\boldsymbol{\nabla}_{\mathbf{k}}u_{n}(\mathbf{k})>$, the Chern
number for the $n$-th quasiparticle band is defined as
\begin{equation}
C_{n}=\frac{1}{2\pi}\iint_{BZ} d^{2}\mathbf{k}\cdot\boldsymbol{\nabla}_{\mathbf{k}}\times\mathbf{A}_{n}(\mathbf{k}).
\end{equation}

When a bulk gap throughout the BZ is present, the TKNN number is just a summation over the Chern numbers of the two negative energy quasiparticle bands, $C=\sum_{n,E_{n}(\mathbf{k}\in BZ)<0}C_{n}=C_{-2}+C_{-1}$. For the present model $C_{\pm2}=0$, thus $C=C_{-1}$. However, such a simple relationship breaks down when the two low-energy quasiparticle bands ($n=\pm1$) overlap in energy. An explicit calculation of $C$ and $C_{n}$ \cite{tknn,fukui05} shows that,
upon crossing the transition point from the bulk gapped phase to the bulk gapless phase, $C$ changes from an integer to a nonintegral value but $C_{n}$ remains quantized (Fig.3(a)). As we tilt the exchange field continuously to the in-plane direction, $C$ continuously approaches zero and $C_{n}$ remains quantized until the exchange field lies \emph{exactly} along the in-plane direction ($\theta=\pi/2$ or $3\pi/2$). By crossing the nodal Weyl superconductivity phase, when the $m_{z}$ component of the exchange field changes sign, the band-wise Chern numbers $C_{n}$ for the two low-energy quasiparticle bands ($n=\pm1$) also change sign (Fig.3(a)). In the Weyl superconductivity phase, the Weyl nodes make the band-wise Chern number of the two low-energy quasiparticle bands ill-defined. And thus no results for $C_{n}$ ($n=\pm1$) are shown in Fig.3(a) for $\theta=\pi/2$ and $3\pi/2$. Therefore, we have shown that while the bulk gapped phase is characterized by both the quantized TKNN number and the quantized band-wise Chern number, the bulk gapless phase with a local gap is characterized by a nonintegral TKNN number and a quantized band-wise Chern number. The Weyl superconductivity phase is a quantum critical point separating two bulk gapless phases with opposite quantized band-wise Chern numbers.

The TKNN number shown in Fig.3(a) can be detected by the thermal Hall effect, which is a transverse thermal current in response to a longitudinal temperature gradient. The formula for the thermal Hall conductivity is \cite{sumiyoshi13,qin11,nomura12}
\begin{equation}
k^{tr}_{xy}=\frac{1}{4\pi T}\int dE E^{2} C(E) \frac{\partial f(E)}{\partial E},
\end{equation}
where $T$ is the temperature, $f(E)=1/(e^{\beta E}+1)$ is the Fermi distribution function ($\beta=1/k_{B}T$, $k_{B}$ is the Boltzmann factor), and $C(E)$ is the TKNN number defined by Eq.(5) but with the energy cutoff (zero) in the definition of $\mathbf{A}(\mathbf{k})$ replaced by $E$. In the low-temperature limit, $k^{tr}_{xy}\simeq-\frac{\pi T}{12}C(E=0)$. The coefficient for the thermal Hall conductivity in the linear-$T$ regime therefore gives a direct measurement of the TKNN number.

\begin{figure}\label{fig3} \centering
\hspace{-5cm} {\textbf{(a)}}\\
\includegraphics[width=6.5cm,height=4.2cm]{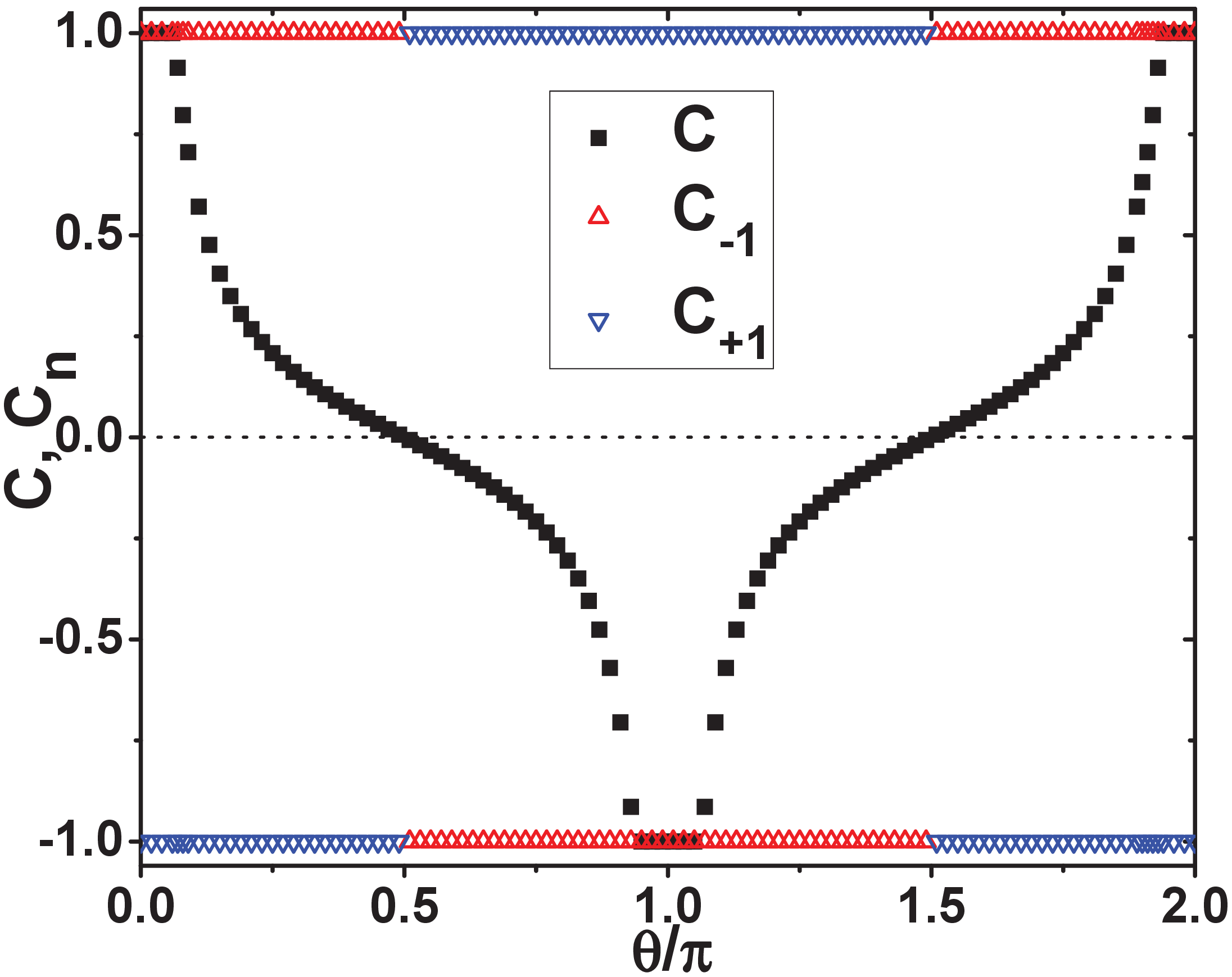} \\ \vspace{-0.05cm}
\hspace{-5cm} {\textbf{(b)}}\\
\includegraphics[width=6.5cm,height=4.2cm]{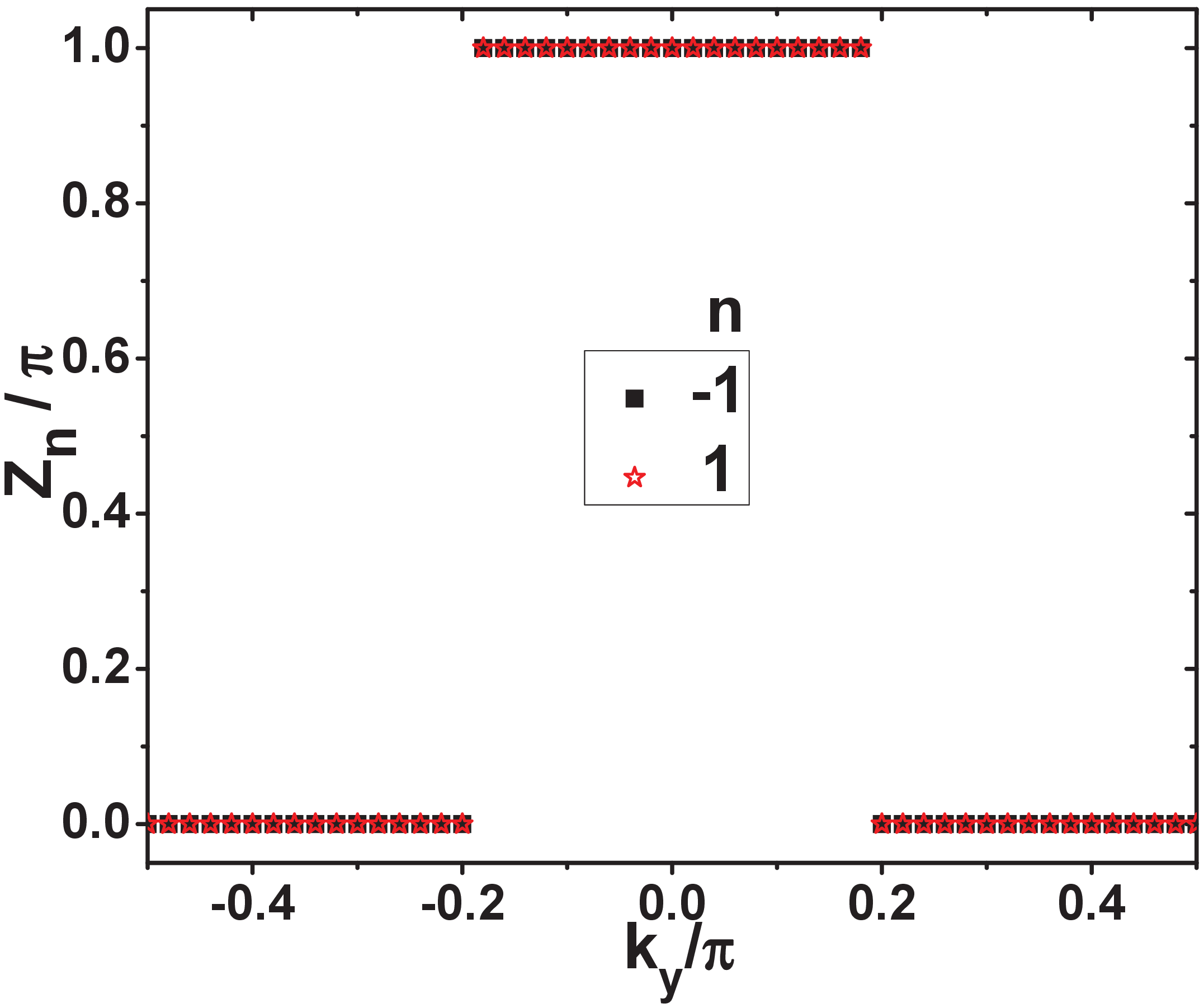}  \\
\vspace{-0.10cm}
\caption{(a)TKNN number ($C$) and the band-wise Chern numbers ($C_{n}$) for the two low-energy quasiparticle bands ($n=\pm1$). (b)The $k_{y}$-resolved Zak phase $Z_{n}$ for the two low-energy quasiparticle bands ($n=\pm1$). The parameters are the same as those for Fig.2.}
\end{figure}

\section{\label{sec:weyl}The 2D Weyl superconductivity phase}

For $\theta=\pi/2$ and $3\pi/2$, the gap between $E_{-1}(\mathbf{k})$ and $E_{1}(\mathbf{k})$ closes at two Weyl nodes with conelike dispersions (See Appendix B for explicit expressions for the low-energy effective models of the Weyl nodes). It is well-known that, accidental band degeneracies in 2D are vanishingly improbable if no symmetry constraint exists.\cite{neumann29,balents11,blount85,hou13} It is thus highly desirable to know whether the Weyl nodes in the present system are ensured by a symmetry. We show that the Weyl nodes and thus the Weyl superconductivity phase are indeed ensured by an emergent symmetry for $\theta=\pi/2$ and $3\pi/2$, which is the mirror reflection symmetry $M_{x}$ ($\tilde{M}_{x}$).

The transformation of $h_{0}(\mathbf{k})$ and $\underline{\Delta}$ under $M_{x}$ are $M^{-1}_{x}h_{0}(\mathbf{k})M_{x}=h_{0}(-k_{x},k_{y})$ and
$M^{-1}_{x}\underline{\Delta}(M^{\text{T}}_{x})^{-1}=\underline{\Delta}$. Turn to the eigenbasis of $h_{0}(\mathbf{k})$, we have $U^{\dagger}(\mathbf{k})h_{0}(\mathbf{k})U(\mathbf{k})=h_{d}(\mathbf{k})$,
where $h_{d}(\mathbf{k})$ is a diagonal matrix storing the eigenenergies of $h_{0}(\mathbf{k})$ and the unitary matrix $U(\mathbf{k})$ stores the eigenvectors in corresponding columns. For the explicit expression of $U(\mathbf{k})$, see Appendix B. The pairing
term becomes $\tilde{\underline{\Delta}}(\mathbf{k})=U^{\dagger}(\mathbf{k})\underline{\Delta}[U^{\dagger}(-\mathbf{k})]^{\text{T}}$. The mirror operation becomes wave vector dependent and is
\begin{equation}
M_{xd}(\mathbf{k})=U^{\dagger}(k_{x},k_{y})M_{x}U(-k_{x},k_{y}).
\end{equation}
$h_{d}(\mathbf{k})$ and $\tilde{\underline{\Delta}}(\mathbf{k})$ now transform as $M^{-1}_{xd}(\mathbf{k})h_{d}(\mathbf{k})M_{xd}(\mathbf{k})=h_{d}(-k_{x},k_{y})$ and
\begin{equation}
M^{-1}_{xd}(\mathbf{k})\tilde{\underline{\Delta}}(\mathbf{k})[M^{\text{T}}_{xd}(-\mathbf{k})]^{-1}=\tilde{\underline{\Delta}}(-k_{x},k_{y}).
\end{equation}

Now, focus on the mirror-invariant lines in the 2D BZ, namely $(k_{x0},k_{y})$ with $k_{x0}=0,\pm\pi$. On these lines, $h_{0}(-k_{x0},k_{y})=h_{0}(k_{x0},k_{y})$. Thus, $M_{x}$ and $h_{0}(k_{x0},k_{y})$ can be diagonalized simultaneously since they commute with each other. Define $U(k_{x0},k_{y})$ as the common eigenvectors of $M_{x}$ and $h_{0}(k_{x0},k_{y})$, clearly we have $U(-k_{x0},k_{y})=U(k_{x0},k_{y})$. From Eq.(8), $M_{xd}(k_{x0},k_{y})$ is diagonal and stores the eigenvalues of $M_{x}$. Since $M_{x}=i\sigma_{1}$ is $\mathbf{k}$-independent,
$M_{xd}(k_{x0},k_{y})$ is also \emph{$\mathbf{k}$-independent} with its two diagonal elements $+i$ and $-i$ acting as a label of the two eigenvalues of $h_{0}(k_{x0},k_{y})$. Explicitly, we have $M_{xd}(k_{x0},k_{y})=i\sigma_{3}$. Substituting it into
Eq.(9), we have
\begin{equation}
i\sigma_{3}\tilde{\underline{\Delta}}(k_{x0},k_{y})i\sigma_{3}=\tilde{\underline{\Delta}}(k_{x0},k_{y}),
\end{equation}
which acts as a constraint on the four elements of the pairing term expressed in the eigenbasis of $h_{0}(k_{x0},k_{y})$. Most importantly, Eq.(10) ensures the vanishing of all intra-band pairing components, namely $\tilde{\underline{\Delta}}_{\alpha\alpha}(k_{x0},k_{y})=0$ ($\alpha=\pm$). The inter-band components, although can be nonzero, only change slightly the value of energy spectrum and do not influence the gap structure of the quasiparticle spectrum. Therefore, we have verified that the Weyl nodes are protected by the mirror symmetry $M_{x}$. From the generality of the above derivation, the protection of superconducting gap nodes along mirror invariant lines
(2D) or planes (3D) is a very robust feature and should be applicable to other relevant systems.

For 3D Weyl semimetal, the existence of the Fermi arc can be understood from the quantized Chern number defined in a reduced 2D subspace of the 3D BZ.\cite{lu15} In a similar sense, the Fermi line on
the edge of the present 2D sample is found to be associated with a quantized topological invariant defined in the reduced 1D subspace of the 2D BZ. This topological invariant is the Zak
phase.\cite{zak89} By taking $k_{y}$ as a parameter, the Zak phase for the $n$-th quasiparticle band is defined as
\begin{equation}
Z_{n}(k_{y})=\int_{-\pi}^{\pi} dk_{x}A^{(x)}_{n}(k_{x},k_{y}).
\end{equation}
For $\theta=\pi/2$ and $3\pi/2$, the mirror symmetry acts effectively as inversion symmetry for the quasi-1D model at each fixed $k_{y}$. The Zak phase in this case is known to take either of two quantized
values, $0$ or $\pi$.\cite{zak89} As shown in Fig.3(b) and in comparison to Fig.2(h), the edge state exists when the Zak phase takes the nontrivial value of $\pi$. Thus, the edge states can be regarded as the end states of the quasi-1D model. The Weyl nodes act as phase
boundaries separating two regions with $0$ and $\pi$ Zak phase. The correspondence between Zak phase and the existence of edge states was also pointed out in graphene.\cite{delplace11}

\section{\label{sec:materials}Experimental considerations}

To observe the predicted topological phase transitions and intermediate Weyl superconductivity phase, it is crucial to have high
quality HM thin films with a single spin-nondegenerate band at the Fermi surface. Fortunately, this has been shown by Chung \emph{et al} to be achievable in several thin film materials, including
atomically thin films of VTe, CrTe, and CrO$_{2}$.\cite{chung11} In addition, a hydrogenated graphene C$_{6}$H$_{1}$ with slight electron doping is predicted recently to be a 2D HM with a
single electron pocket around the $\Gamma$ point.\cite{lu16} This ideal 2D material provides a perfect playground to realize the predictions in the present work. Besides the materials mentioned above, it is
still an open question whether a simple Fermi surface with a single spin-polarized band can be obtained in other materials. The possibility could be
manganites\cite{dagotto01,salamon01}, double perovskites like Sr$_{2}$FeMoO$_{6}$\cite{erten11}, and electron-doped HgCr$_{2}$Se$_{4}$\cite{guan15}.

Another important requirement to be satisfied is the possibility to tune the direction of the spontaneous magnetization, $\theta$. In principle, this can be achieved by two kinds of methods depending
on the specific material. If the magnetic anisotropy of the material is small, $\theta$ can be tuned over a large range with an external magnetic field above the superconducting transition temperature. In this case, all the phase transitions
together with the intermediate Weyl superconductivity phase can be observed. On the other hand, for a material with large magnetic anisotropy, $\theta$ can still be tuned by controlling the direction
along which the thin film is cut from the bulk HM. Though only some particular $\theta$ can be attained in this case, some typical examples of the different phases can still be observed. In
particular, since the easy axis of the magnetization is usually along a high symmetry direction, the condition for the realization of the Weyl superconductivity phase can still be fulfilled.

Finally, since the heterostructure consisting of a ferromagnet and a superconductor is a standard element in the superconducting spintronics, a great deal of experimental experiences have been accumulated.\cite{eschrig03,buzdin05,bergeret05,eschrig11,linder15,kolenda16} Besides, the issue of choosing a proper substrate to generate a large RSOC is also a mature field.\cite{nitta97,marchenko12,zhou14} Thus, our predictions should have good prospect to be realized by experiments in the near future.

\section{\label{sec:summary}Summary}

To summarize, we have found topological phase transitions in a HM/$s$SC heterostructure as the orientation of the magnetization varies. While the transition between a fully gapped phase and a gapless phase can be distinguished by a change in the TKNN number, the transition between the gapless phase with a local gap and a Weyl superconductivity phase is captured by a band-wise Chern number. The protection of the Weyl superconductivity phase by an emergent mirror symmetry is established. The position of the Weyl nodes and the existence of Fermi line edge states are captured by the Zak phase defined in the 1D subspace of the 2D BZ. In order for our predictions to be confirmed by future experiments, it is important to seek the materials to form the HM/$s$SC heterstructure.

\begin{acknowledgments}
This work was supported by the Texas Center for Superconductivity at the University of Houston and the Robert A. Welch Foundation (Grant No. E-1146). L.H. would also like to acknowledge the support from the China Scholarship Council. The numerical calculations were performed at the Center of Advanced Computing and Data Systems at the University of Houston.
\end{acknowledgments}\index{}

\begin{appendix}

\section{Pfaffian Z$_{2}$ topological invariant}

An important symmetry of our model is the particle-hole symmetry, which is expressed as $\Xi^{-1}h(\mathbf{k})\Xi=-h(-\mathbf{k})$. The particle-hole operator is represented as $\Xi=\tau_{1}\sigma_{0}K=\Lambda K$, where $K$ denotes complex conjugation. As was shown by Ghosh \emph{et al}, a $Z_{2}$ topological invariant can be defined from the particle-hole symmetry of the model.\cite{ghosh10} Namely, at the particle-hole-invariant momenta $\mathbf{K}_{i}$, an antisymmetric matrix can be defined as $W(\mathbf{k})=h(\mathbf{K}_{i})\Lambda$. In terms of the Pfaffian of $W(\mathbf{k})$, a number that takes on discrete values of $\pm1$ is defined as
\begin{equation}
Q[h(\mathbf{K}_{i})]=\text{sgn}\{i^{n}Pf[W(\mathbf{K}_{i})]\},
\end{equation}
where $n=2$ is half of the rank of $h(\mathbf{k})$. Since $Pf[W(\mathbf{K}_{i})]^{2}=\text{det}[h(\mathbf{K}_{i})\Lambda]=\text{det}[h(\mathbf{K}_{i})]$, $Q[h(\mathbf{K}_{i})]$ changes sign only when $h(\mathbf{K}_{i})$ has a zero eigenvalue. Therefore, $Q[h(\mathbf{K}_{i})]$ is invariant once there is a gap in the spectrum of $h(\mathbf{K}_{i})$. It was further illustrated by Ghosh \emph{et al} that, when the Chern number is well defined, its parity is directly related to the product of the four invariants at the four $\mathbf{K}_{i}$ ($\mathbf{K}_{1}=(0,0)$, $\mathbf{K}_{2}=(\pi,0)$, $\mathbf{K}_{3}=(0,\pi)$, $\mathbf{K}_{4}=(\pi,\pi)$). The $Z_{2}$ invariant related to all four particle-hole-invariant momenta is defined as
\begin{equation}
P=\frac{Q[h(0,0)]Q[h(\pi,\pi)]}{Q[h(\pi,0)]Q[h(0,\pi)]}.
\end{equation}
For our present model defined as Eq.(2) of the main text, we have
\begin{equation}
W(\mathbf{K}_{i})=\epsilon_{\mathbf{K}_{i}}i\tau_{2}\sigma_{0}+m_{x}i\tau_{2}\sigma_{1}+m_{z}i\tau_{2}\sigma_{3}+\Delta_{0} i\tau_{3}\sigma_{2}.
\end{equation}
The Pfaffian and the $Z_{2}$ invariant for $\mathbf{K}_{i}$ are
\begin{equation}
\text{Pf}[W(\mathbf{K}_{i})]=m_{x}^{2}+m_{z}^{2}-\Delta_{0}^{2}-\epsilon^{2}_{\mathbf{K}_{i}}=m^{2}-\Delta_{0}^{2}-\epsilon^{2}_{\mathbf{K}_{i}},
\end{equation}
and
\begin{equation}
Q[h(\mathbf{K}_{i})]=\text{sgn}\{\epsilon^{2}_{\mathbf{K}_{i}}+\Delta_{0}^{2}-m^{2}\}.
\end{equation}
Recalling the definition $\epsilon_{\mathbf{k}}=-2t(\cos k_{x}+\cos k_{y})-\mu$, and our assumption that $\mu\simeq-4t$, we have
\begin{equation}
P=Q[h(\mathbf{K}_{1})]=\text{sgn}\{\Delta_{0}^{2}+\epsilon^{2}_{\mathbf{K}_{1}}-m^{2}\}.
\end{equation}
An important feature of the above expression is that it depends only on the magnitude $m$ of the exchange field and does not see the difference of phases induced by the changes in $\theta$.

\section{low-energy effective model}
Because the pairing amplitude is usually much smaller than other important energy scales in the problem, such as the chemical potential and the exchange field in the present HM, only the low-energy spin-polarized band which crosses the chemical potential (thus contributing to the Fermi surface) is important in the analysis of the physical properties. A simplified approach of studying the low-energy physics is to project the original model containing information both of the high-energy band and of the low-energy band to an effective model retaining only information of the low-energy band.\cite{yip13,hao15} In this section, we implement this reduction of model and show in some detail several physical quantities that can be studied in terms of this approach.

We first repeat the definition of the model. Denoting the basis vector as $\phi^{\dagger}_{\mathbf{k}}=[d^{\dagger}_{\mathbf{k}\uparrow},d^{\dagger}_{\mathbf{k}\downarrow}]$, the model Hamiltonian for the HM thin film with a RSOC term induced by the formation of the heterostructure is $\hat{H}_{0}=\sum_{\mathbf{k}}\phi^{\dagger}_{\mathbf{k}}h_{0}(\mathbf{k})\phi_{\mathbf{k}}$, where
\begin{equation}
h_{0}(\mathbf{k})=\epsilon_{\mathbf{k}}\sigma_{0}+m_{x}\sigma_{1}+m_{z}\sigma_{3}+\lambda(\sin k_{x}\sigma_{2}-\sin k_{y}\sigma_{1}).
\end{equation}
$\epsilon_{\mathbf{k}}=-2t(\cos k_{x}+\cos k_{y})-\mu$, $m_{x}=m\sin\theta$, $m_{z}=m\cos\theta$. Definition of other parameters are as explained in the main text. The proximity-induced superconductivity in the HM arising from coupling with an $s$SC is described by $\hat{H}_{p}=\frac{1}{2}\sum_{\mathbf{k}}\phi^{\dagger}_{\mathbf{k}}\underline{\Delta}(\mathbf{k})\phi^{\dagger}_{-\mathbf{k}}+\text{H.c.}$, where
$\underline{\Delta}(\mathbf{k})=\Delta_{0}(\mathbf{k})i\sigma_{2}$. For the sake of simplicity, we neglect the wave vector dependence of the pairing amplitude by taking $\Delta_{0}(\mathbf{k})=\Delta_{0}$ and so $\underline{\Delta}(\mathbf{k})=\underline{\Delta}$. As usually is the case, we assume $\Delta_{0}$ much smaller than the leading energy scales ($t$, $m$, and $\mu+4t$) in the problem.

\begin{figure}[!htb]\label{fig4}
\centering
\hspace{-5cm} {\textbf{(a)}}\\
\includegraphics[width=6.5cm,height=4.2cm]{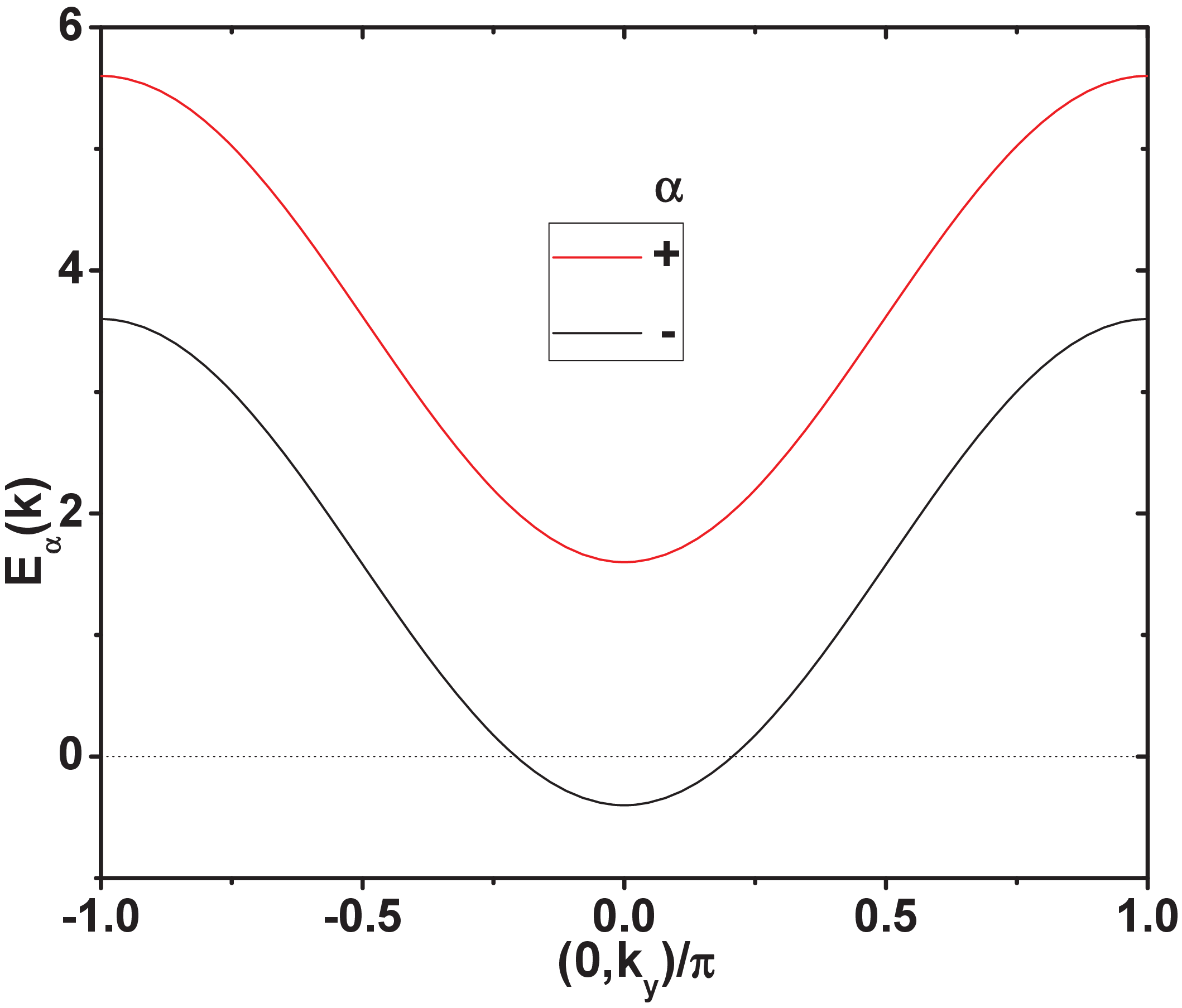} \\ \vspace{-0.05cm}
\hspace{-5cm} {\textbf{(b)}}\\
\includegraphics[width=6.5cm,height=4.2cm]{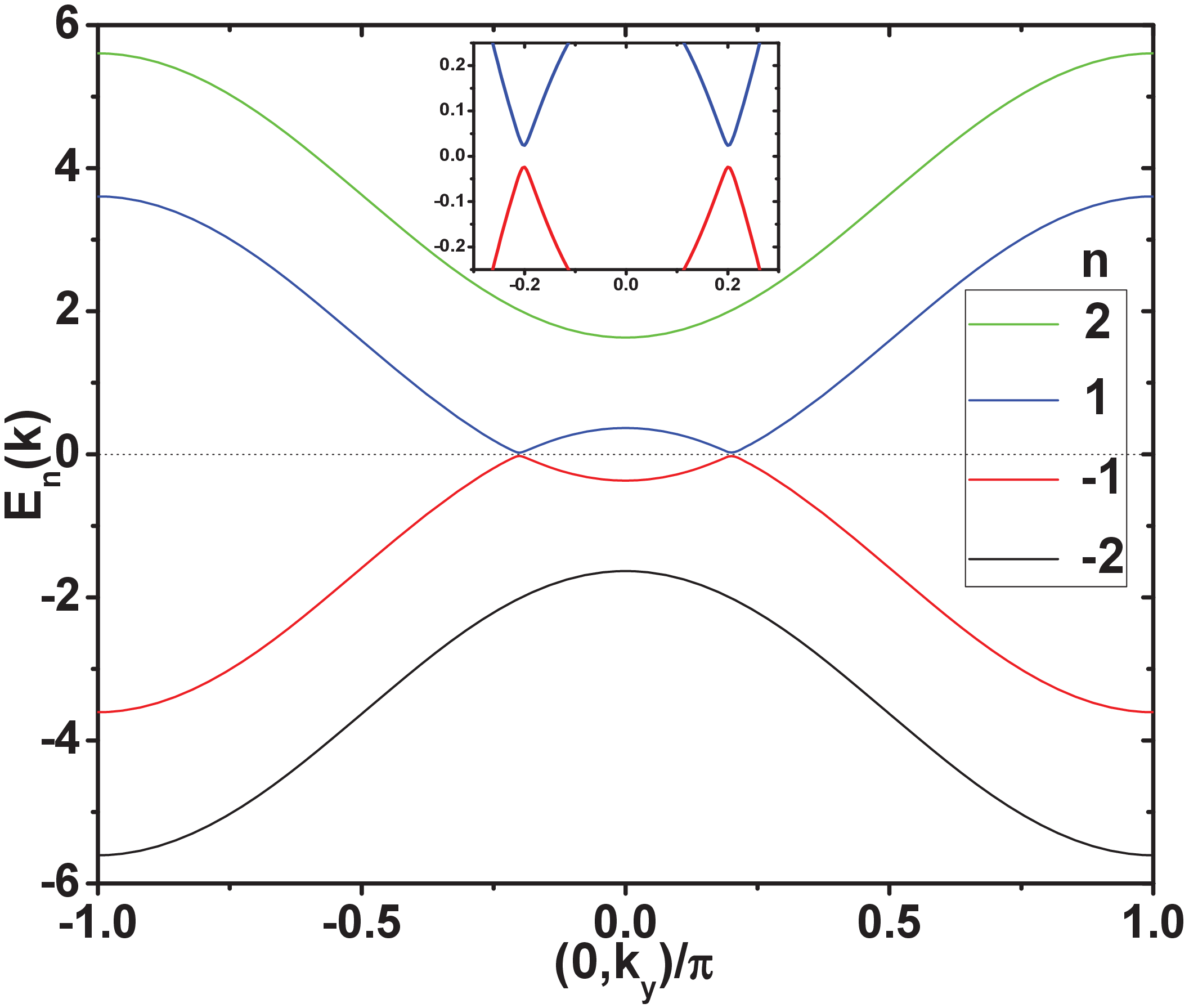}  \\
\caption{(a) $E_{\alpha}(\mathbf{k})$ ($\alpha=\pm$) for $m=t>0$, $\lambda=0.2t$, $\mu=-4.6t$, and $\theta=0$. (b) $E_{n}(\mathbf{k})$ ($n=-2$, $-1$, $1$, $2$) for $m=t>0$, $\lambda=\Delta_{0}=0.2t$, $\mu=-4.6t$, and $\theta=0$. As is shown in the inset of (b), there is a small gap of the size of about 0.05 eV between $E_{-1}(\mathbf{k})$ and $E_{1}(\mathbf{k})$. The energies are in unit of $t$. The horizontal dotted lines mark the positions of the chemical potential.}
\end{figure}

$h_{0}(\mathbf{k})$ can be diagonalized by a unitary transformation $U^{\dagger}(\mathbf{k})h_{0}(\mathbf{k})U(\mathbf{k})=h_{d}(\mathbf{k})$, where $h_{d}(\mathbf{k})$ is a diagonal matrix with the diagonal elements storing the two eigenvalues of $h_{0}(\mathbf{k})$, and $U(\mathbf{k})$ is a unitary matrix containing the eigenvectors of $h_{0}(\mathbf{k})$ in corresponding columns. The two eigenenergies of $h_{0}(\mathbf{k})$ are
\begin{eqnarray}
E_{\alpha}(\mathbf{k})&=&\epsilon_{\mathbf{k}}+\alpha\sqrt{m_{z}^{2}+(m_x-\lambda\sin k_{y})^{2}+\lambda^{2}\sin^{2} k_{x}} \notag \\
&=&\epsilon_{\mathbf{k}}+\alpha\tilde{E}(\mathbf{k}),
\end{eqnarray}
where $\alpha=\pm$. Notice that $E_{\alpha}(\mathbf{k})$ ($\alpha=\pm$) are the electronic energy bands in the normal state (i.e., without the proximity-induced pairing term) and should be distinguished from the quasiparticle bands $E_{n}(\mathbf{k})$ ($n=$-2, -1, 1, 2) defined as eigenvalues of Eq.(2) in the main text. As shown in Fig.4(a) and Fig.4(b) are plots for $E_{\alpha}(\mathbf{k})$ and $E_{n}(\mathbf{k})$ along $(k_{x}=0,k_{y})$ and for $\theta=0$. In all other figures of this paper, only the low-energy parts of the energy spectrum are shown. $U(\mathbf{k})$ is written as
\begin{equation}
U(\mathbf{k})=\begin{pmatrix} u_{+}(\mathbf{k}) & u_{-}(\mathbf{k}) \\
v_{+}(\mathbf{k}) & v_{-}(\mathbf{k})
\end{pmatrix},
\end{equation}
where we choose
\begin{equation}
\begin{pmatrix} u_{\alpha}(\mathbf{k}) \\
v_{\alpha}(\mathbf{k})\end{pmatrix}=\frac{1}{D_{\alpha}(\mathbf{k})}\begin{pmatrix}
m_{x}-\lambda\sin k_{y}-i\lambda\sin k_{x} \\
\alpha\tilde{E}(\mathbf{k})-m_{z} \end{pmatrix},
\end{equation}
where $\alpha=\pm$ and $D_{\alpha}(\mathbf{k})=\sqrt{2\tilde{E}(\mathbf{k})(\tilde{E}(\mathbf{k})-\alpha m_{z})}$. Clearly, for $\lambda\ne0$, up-spin and down-spin states are mixed in a nontrivial manner, and therefore eigenstates in each band no longer point to the same direction in the spin subspace. In the eigenbasis, the pairing term is rewritten as
\begin{equation}
\tilde{\underline{\Delta}}(\mathbf{k})=U^{\dagger}(\mathbf{k})\underline{\Delta}U^{\ast}(-\mathbf{k}).
\end{equation}
Since only the $E_{-}(\mathbf{k})$ band contributes to the low-energy properties, we can retain only this band and the pairing term within this band to construct the low-energy effective model. In the reduced Nambu space defined by the basis $\tilde{\varphi}^{\dagger}_{\mathbf{k}}=[d^{\dagger}_{\mathbf{k},-},d_{-\mathbf{k},-}]$, the low-energy effective model is written as
\begin{eqnarray}
\tilde{H}&=&\frac{1}{2}\sum\limits_{\mathbf{k}}\tilde{\varphi}^{\dagger}_{\mathbf{k}}\begin{pmatrix} E_{-}(\mathbf{k}) &  \tilde{\Delta}_{--}(\mathbf{k}) \\
\tilde{\Delta}^{\ast}_{--}(\mathbf{k}) & -E_{-}(-\mathbf{k})
\end{pmatrix}\tilde{\varphi}_{\mathbf{k}}   \notag \\
&=&\frac{1}{2}\sum\limits_{\mathbf{k}}\tilde{\varphi}^{\dagger}_{\mathbf{k}} \tilde{h}(\mathbf{k}) \tilde{\varphi}_{\mathbf{k}}.
\end{eqnarray}
The intra-band pairing amplitude is
\begin{equation}
\tilde{\Delta}_{--}(\mathbf{k})=\Delta_{0}[u^{\ast}_{-}(\mathbf{k})v^{\ast}_{-}(-\mathbf{k})-u^{\ast}_{-}(-\mathbf{k})v^{\ast}_{-}(\mathbf{k})].
\end{equation}
For $\theta=0$, we have
\begin{equation}
\tilde{\Delta}_{--}(\mathbf{k})=-i\Delta_{0}\frac{\lambda}{\tilde{E}(\mathbf{k})}(\sin k_{x}+i\sin k_{y}),
\end{equation}
which indicates the formation of chiral $p_{x}+ip_{y}$ pairing on the Fermi surface.\cite{chung11} The effective pairing interaction vanishes in the limit of $\lambda=0$, even if $\Delta_{0}\ne0$. The explicit expression of $\tilde{\Delta}_{--}(\mathbf{k})$ for general $\theta$ is cumbersome. However, a universal property is that $\tilde{\Delta}_{--}(\mathbf{k})$ vanishes for $\lambda=0$. We will analyze in what follows two special cases, from which we can extract several important quantities of the system.

\subsection{The critical angle $\theta_{c}$}

The first quantity we would like to determine is the critical angle $\theta_{c}$ marking the first phase transition from the phase with a bulk gap to the phase without a bulk gap but has a ubiquitous local gap in the momentum space. The superconducting gap opens along the Fermi circle determined by $E_{-}(\mathbf{k})=-E_{-}(-\mathbf{k})$. Since $E_{-}(\mathbf{k})$ is symmetrical in $k_{x}$, the transition is driven by the qualitative change of the quasiparticle spectrum along the $k_y$ direction. Thus, we focus on the line $(k_x=0,k_y)$. The two Fermi points in this direction that lie on the Fermi circle are determined by
\begin{equation}
\epsilon_{0,k_y}=\frac{1}{2}\sum\limits_{\alpha=\pm}\sqrt{m_{z}^{2}+(m_{x}+\alpha\lambda\sin k_{y})^{2}}.
\end{equation}
Focusing on the limit of $|\lambda|/m\ll1$, we can make a Taylor expansion of the square roots in Eq.(B9) over $\lambda$ and keep the leading order terms and gets
\begin{equation}
-2t\cos k_{y}-(\mu+2t)=m(1+\frac{\lambda^{2}}{2m^{2}}\sin^{2}k_{y}).
\end{equation}
Among the two solutions to $\cos k_{y}$ from the above equation,
\begin{equation}
\cos k_{y}=\frac{|2mt|\pm\sqrt{4m^{2}t^{2}+\lambda^{2}(2m^{2}+2m(\mu+2t)+\lambda^{2})}}{\text{sgn}(mt)\lambda^{2}},
\end{equation}
the physical one is clearly the one corresponding to the minus sign. The equation for the Fermi points can be further simplified to
\begin{equation}
\cos k_{y0}\simeq-\frac{m+\mu+2t}{2t}.
\end{equation}
Take $k_{y0}=\arccos[-\frac{m+\mu+2t}{2t}]$. The energy gaps are thus introduced at $(0,k_{y0})$ and $(0,-k_{y0})$ with energies $E_{-}(0,k_{y0})$ and $E_{-}(0,-k_{y0})$, respectively. The corresponding pairing amplitudes are $\tilde{\Delta}_{--}(0,k_{y0})$ and $\tilde{\Delta}_{--}(0,-k_{y0})$. The transition point $\theta_{c}$ is arrived at when the composite gap amplitude
\begin{equation}
\Delta(k_{y0})=|\tilde{\Delta}_{--}(0,k_{y0})|+|\tilde{\Delta}_{--}(0,-k_{y0})|
\end{equation}
equals the energy difference between the two Fermi points
\begin{equation}
E(k_{y0})=|E_{-}(0,k_{y0})-E_{-}(0,-k_{y0})|.
\end{equation}
In the limit of $|\lambda|/m\ll1$ (together with the assumption of $|\Delta_{0}|/m\ll1$ made in deriving the low-energy effective model), the condition $\Delta(k_{y0})=E(k_{y0})$ gives
\begin{equation}
|\tan\theta_{c}|=\frac{|\Delta_{0}|}{m}.
\end{equation}
Therefore we have reproduced Eq.(3) in the main text. Note that Eq.(B15) gives the approximate values of all the four critical angles (in the whole range of $\theta\in[0,2\pi)$) which separate the bulk gapped phase and the bulk gapless phase with a ubiquitous local gap.

\subsection{The Weyl nodes and the effective models close to the Weyl nodes}

In the main text, we have proved that the emergent mirror symmetry for $\theta=\pi/2$ (and also for $\theta=3\pi/2$) ensures the presence of Weyl nodes along the mirror invariant lines which cross the Fermi circle. Here, we reconfirm this conclusion from the explicit expression of the low-energy effective model.

For $\theta=\pi/2$, $m_{x}=m$ and $m_{z}=0$. The effective pairing amplitude has a simple form
\begin{eqnarray}
\tilde{\Delta}_{--}(\mathbf{k})&=&\frac{\Delta_{0}}{2}[\lambda(\sin k_{y}-i\sin k_{x})(\frac{1}{M_{+}(\mathbf{k})}+\frac{1}{M_{-}(\mathbf{k})})  \notag \\
&&+m(\frac{1}{M_{+}(\mathbf{k})}-\frac{1}{M_{-}(\mathbf{k})})],
\end{eqnarray}
where $M_{\alpha}(\mathbf{k})=\sqrt{(m+\alpha\lambda\sin k_{y})^{2}+\lambda^{2}\sin^{2}k_{x}}$ ($\alpha=+$ or $-$). On all mirror-invariant lines along which $\sin k_{x}=0$, the pairing amplitude $\tilde{\Delta}_{--}(\sin k_{x}=0,k_{y})=0$. Thus, once the Fermi circle crosses with one or several mirror-invariant line, it will give one or several pairs of Weyl nodes and result in a Weyl superconductivity phase. For our model and assumption on the parameters ($\mu\sim-4t$), the relevant mirror-invariant line crossing the Fermi circle is along $(0, k_{y})$. The Weyl points are determined by the condition of $E_{-}(0,k_{y0})=-E_{-}(0,-k_{y0})$, which gives approximately for $|\lambda|/m\ll1$
\begin{equation}
\cos k_{y0}\simeq-\frac{m+\mu+2t}{2t}.
\end{equation}
For the parameters that we focus on, the above equation gives us a pair of solutions which correspond to the two Weyl nodes at $k_{x}=0$ and
\begin{equation}
k_{y0}\simeq\pm\arccos[-\frac{m+\mu+2t}{2t}].
\end{equation}

Introducing the relative momenta $q_{x}$ and $q_{y}$ close to the Weyl node $(0,k_{y0})$, the low-energy effective model can be written in terms of $q_{x}$ and $q_{y}$ by making series expansions to the terms in $\tilde{h}(\mathbf{k})$ and retaining the leading order terms. The result turns out to be
\begin{equation}
\tilde{h}(q_{x},k_{y0}+q_{y})\simeq \sum\limits_{i=0}^{3}d_{i}(q_{x},q_{y};k_{y0})\tau_{i},
\end{equation}
where
\begin{eqnarray}
d_{0}(q_{x},q_{y};k_{y0})&=&E_{-}(0,k_{y0})+(\lambda\cos k_{y0})q_{y}    \notag \\
&=&E_{-}(0,k_{y0})+v_{0}q_{y},
\end{eqnarray}
\begin{equation}
d_{1}(q_{x},q_{y};k_{y0})=0,
\end{equation}
\begin{equation}
d_{2}(q_{x},q_{y};k_{y0})=\frac{\lambda m\Delta_{0}}{m^{2}-\lambda^{2}\sin^{2}k_{y0}}q_{x}=v_{2}q_{x},
\end{equation}
and
\begin{equation}
d_{3}(q_{x},q_{y};k_{y0})=(2t\sin k_{y0})q_{y}=v_{3}q_{y}.
\end{equation}
In the presence of the $d_{0}$ term, on one hand the two Weyl nodes have different energies, on the other hand the two cones are tilted along the $k_{y}$ direction, which are both clear from Fig.2(d) of the main text. $d_{1}(q_{x},q_{y};k_{y0})=0$ is consistent with our conclusion in the main text that the effective pairing amplitude vanishes along the $(0,k_{y})$ direction, which is ensured by the mirror reflection symmetry. The coefficient of the $d_{3}$ term, $2t\sin k_{y0}=v_{3}$, has opposite sign for the two nodes. Therefore, we can define the chirality for the two Weyl nodes as $c=\text{sgn}(v_{2}v_{3})$, which takes the value of $+1$ and $-1$ for the two nodes.

\section{experimental features of the edge states}

In Figs.2(e)-e(h) of the main text, we have shown the energy spectra for strips of the system at several typical values of $\theta$. The chiral edge states traversing the (local) gap are seen clearly. On the other hand, the experimentally relevant quantity related to the edge states are the spectral function for the edge layers, rather than the full quasiparticle spectrum. In this section, we give more numerical results on the experimentally relevant spectroscopic properties of the chiral Majorana edge states. These include the spectral functions and the local density of states of the two edge layers. The spectral function is defined as imaginary part of the retarded Green's function for states on the edge layers. The integration of the spectral function over the one-dimensional edge BZ ($-\pi\le k_{y}<\pi$) then gives the density of states (DOS).

To get the Green's functions for the two edge layers of a strip with two edges parallel (perpendicular) to the $y$ ($x$) axis, we bring the $x$ coordinate of the model to the real space.  This is achieved by making a partial Fourier transformation to Eq.(2) of the main text in terms of
\begin{equation}
d_{\mathbf{k}\sigma}=\frac{1}{\sqrt{N_x}}\sum\limits_{n_{x}}d_{n_{x}k_{y}\sigma}e^{-ik_{x}n_{x}},
\end{equation}
where $\sigma$ is the spin label, $N_{x}$ is the number of unit cells (layers) of the strip along the $x$ direction, $n_{x}$ is a label for the layers along $x$ and takes the value from $1$ to $N_{x}$. The lattice constant has been taken as the length unit. In terms of the Nambu basis defined in the mixed $(n_{x},k_{y})$ space, $\varphi^{\dagger}_{n_{x}k_{y}}=[\phi^{\dagger}_{n_{x}k_{y}},\phi^{\text{T}}_{n_{x},-k_{y}}]$, the model is no longer diagonal in $n_{x}$ and takes the form
\begin{equation}
\hat{H}=\frac{1}{2}\sum_{n_{x}k_{y}}\varphi^{\dagger}_{n_{x}k_{y}}[h'(k_{y})\varphi_{n_{x}k_{y}}+h_{+}\varphi_{n_{x}+1,k_{y}}+h_{-}\varphi_{n_{x}-1,k_{y}}],
\end{equation}
where
\begin{eqnarray}
h'(k_{y})&=&\tilde{\epsilon}_{\mathbf{k}}\tau_{3}\sigma_{0}+m_{x}\tau_{3}\sigma_{1}+m_{z}\tau_{3}\sigma_{3}    \notag \\
&&-\lambda\sin k_{y}\tau_{0}\sigma_{1}-\Delta_{0}\tau_{2}\sigma_{2},
\end{eqnarray}
\begin{equation}
h_{+}=-t\tau_{3}\sigma_{0}+\frac{\lambda}{2i}\tau_{3}\sigma_{2},
\end{equation}
and
\begin{equation}
h_{-}=h^{\dagger}_{+}=-t\tau_{3}\sigma_{0}-\frac{\lambda}{2i}\tau_{3}\sigma_{2}.
\end{equation}
$\tilde{\epsilon}_{\mathbf{k}}=-2t\cos k_{y}-\mu$. The retarded Green's function for an isolated layer is defined as
\begin{equation}
g(k_{y},\omega)=[\omega+i\eta-h'(k_{y})]^{-1},
\end{equation}
where $\eta$ is the positive infinitesimal and will be taken as a small finite positive number (i.e., $\eta=10^{-5}t$) in actual calculations. Consider a strip that is wide enough (i.e., $N_{x}\rightarrow\infty$) so that finite size effect is absent. Denote the retarded Green's function for the two edge layers with $n_{x}=1$ and $n_{x}=N_{x}$ as $G_{L}$ and $G_{R}$ respectively. $G_{L}(k_{y},\omega)$ and $G_{R}(k_{y},\omega)$ are obtained iteratively in terms of the following formula \cite{hao11}
\begin{equation}
G^{(m)}_{L}(k_{y},\omega)=[g^{-1}(k_{y},\omega)-h_{+}G^{(m-1)}_{L}(k_{y},\omega)h_{-}]^{-1},
\end{equation}
and
\begin{equation}
G^{(m)}_{R}(k_{y},\omega)=[g^{-1}(k_{y},\omega)-h_{-}G^{(m-1)}_{R}(k_{y},\omega)h_{+}]^{-1}.
\end{equation}
$m\ge1$ is the label for the number of iterations that has been performed. The iterative calculation starts with $G^{(0)}_{L}(k_{y},\omega)=G^{(0)}_{R}(k_{y},\omega)=g(k_{y},\omega)$ and ends when the differences between every matrix element of $G^{(m)}_{L(R)}(k_{y},\omega)$ and that of $G^{(m-1)}_{L(R)}(k_{y},\omega)$ is smaller than a certain precision set by hand. The converged $G^{(m)}_{L}(k_{y},\omega)$ and $G^{(m)}_{R}(k_{y},\omega)$ are then taken as approximations to $G_{L}(k_{y},\omega)$ and $G_{R}(k_{y},\omega)$. The spectral functions for states on the two edges are then obtained from
\begin{equation}
A_{L}(k_{y},\omega)=-\frac{1}{\pi}\sum\limits_{i=1}^{2}\text{Im}[G_{L}(k_{y},\omega)]_{ii},
\end{equation}
and
\begin{equation}
A_{R}(k_{y},\omega)=-\frac{1}{\pi}\sum\limits_{i=1}^{2}\text{Im}[G_{R}(k_{y},\omega)]_{ii},
\end{equation}
where $\text{Im}$ means taking the imaginary part of the specified diagonal matrix element of the Green's function. Finally, the density of states (DOS) for the two edges are obtained by summing over states in the edge BZ
\begin{equation}
\rho_{L}(\omega)=\frac{1}{N_{y}}\sum\limits_{k_{y}}A_{L}(k_{y},\omega),
\end{equation}
and
\begin{equation}
\rho_{R}(\omega)=\frac{1}{N_{y}}\sum\limits_{k_{y}}A_{R}(k_{y},\omega),
\end{equation}
where $N_{y}$ is the number of unit cells in the sample along the $y$ direction, which is also the number of $k_{y}$ in the edge BZ.

\begin{figure}\label{fig5} \centering
\hspace{-2.95cm} {\textbf{(a)}} \hspace{3.8cm}{\textbf{(b)}}\\
\hspace{0cm}\includegraphics[width=4.2cm,height=3.50cm]{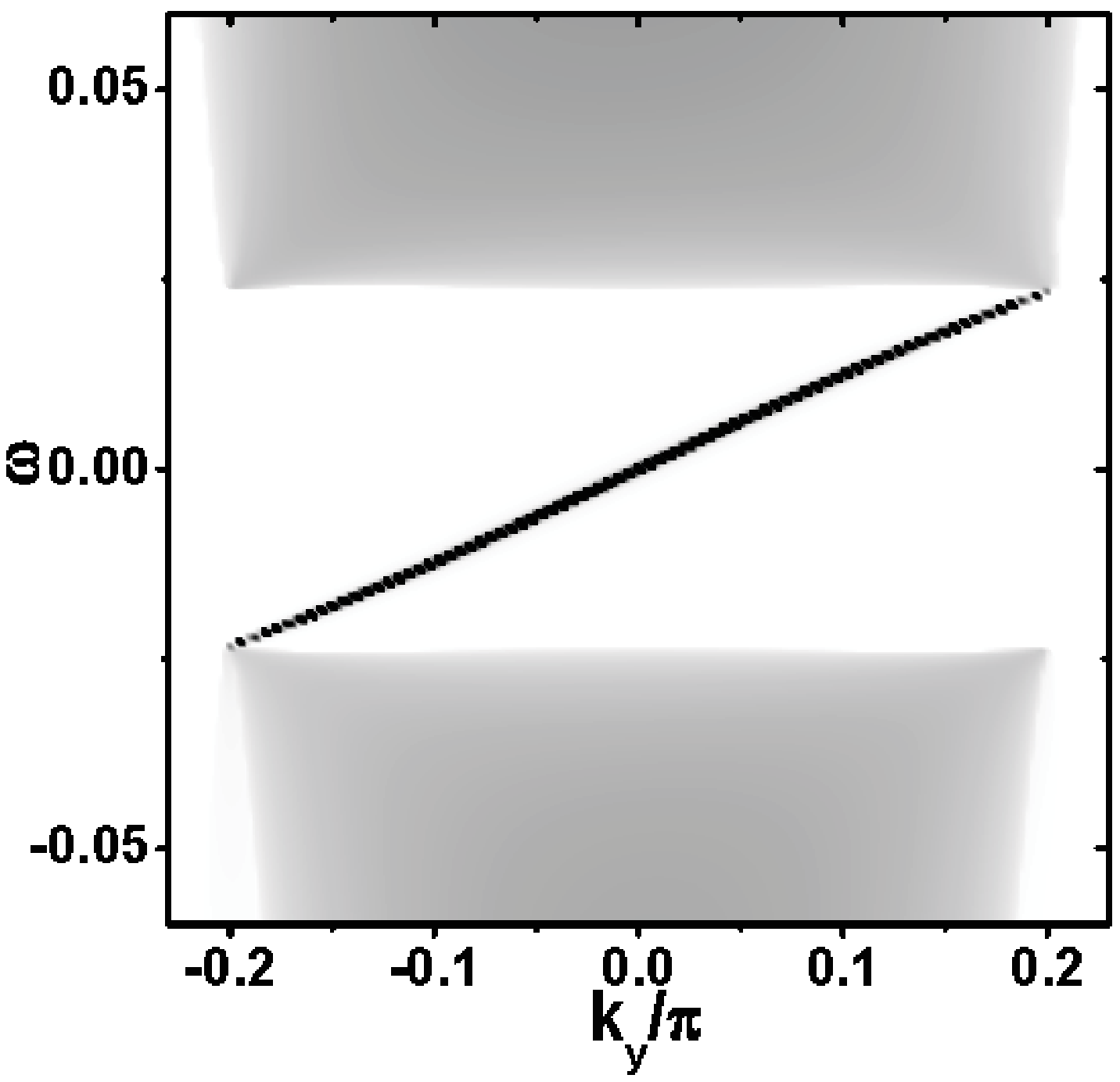}
\includegraphics[width=4.2cm,height=3.50cm]{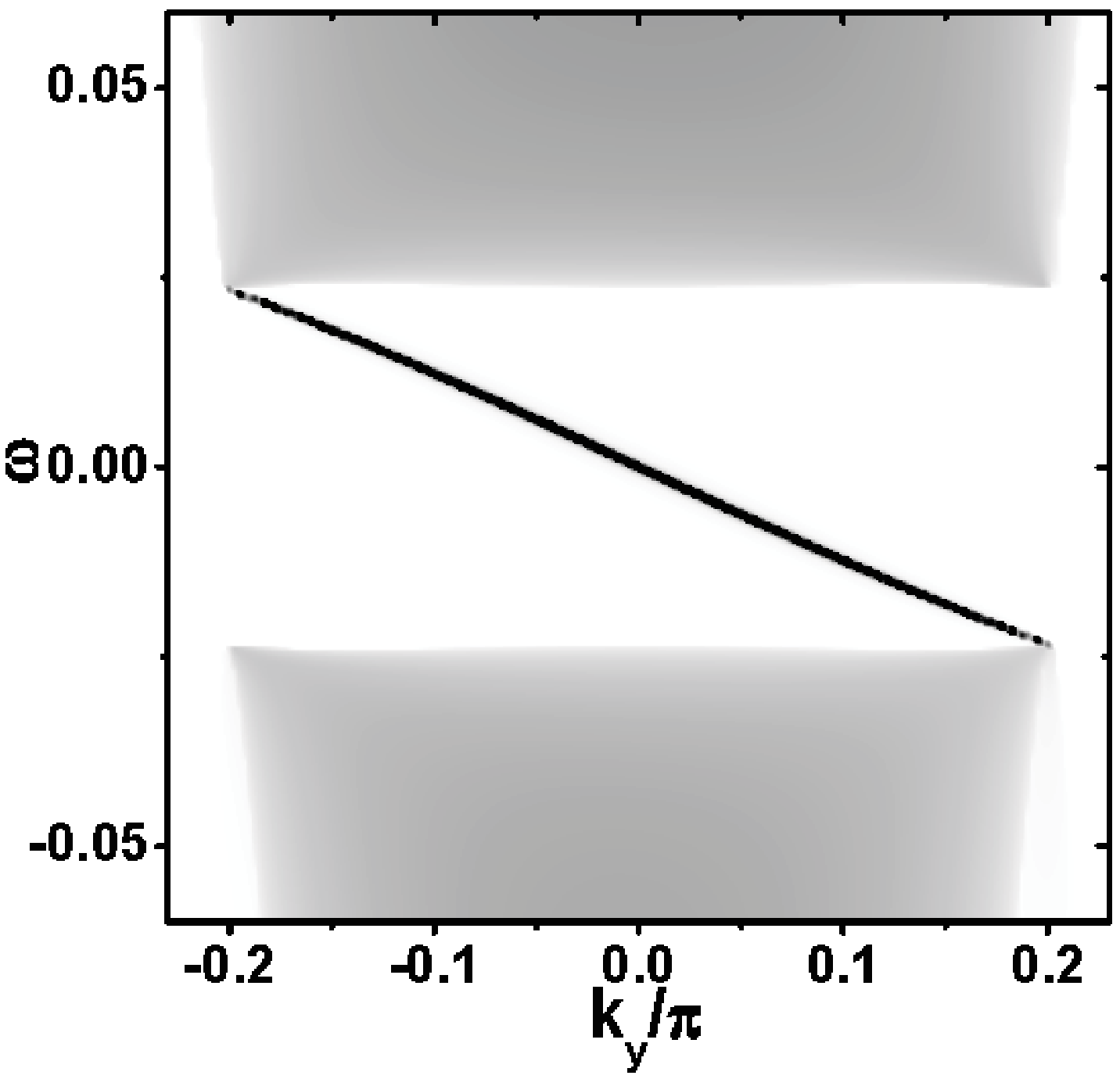} \vspace{-0.10cm} \\
\hspace{-2.95cm} {\textbf{(c)}} \hspace{3.8cm}{\textbf{(d)}}\\
\hspace{0cm}\includegraphics[width=4.2cm,height=3.50cm]{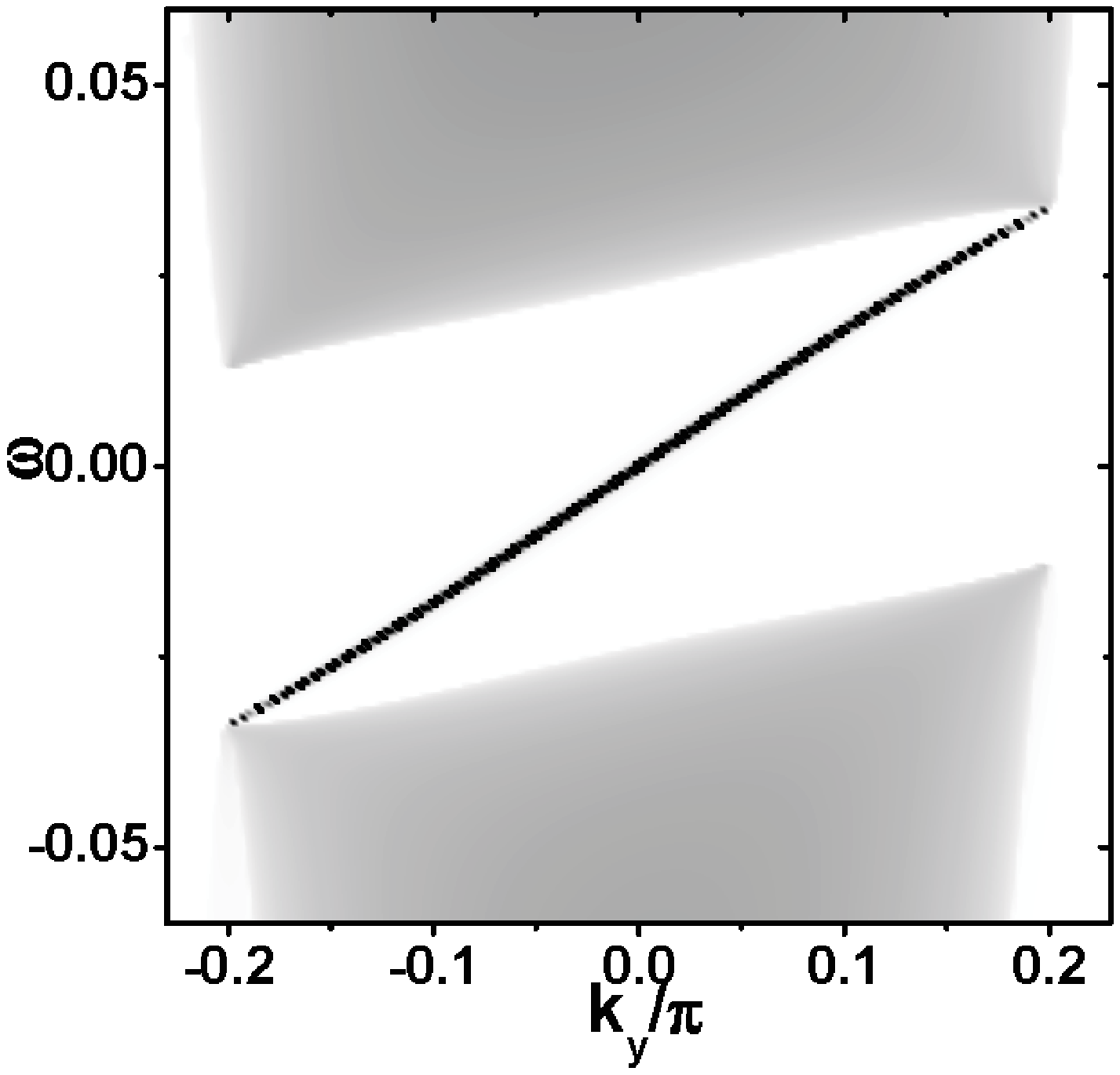}
\includegraphics[width=4.2cm,height=3.50cm]{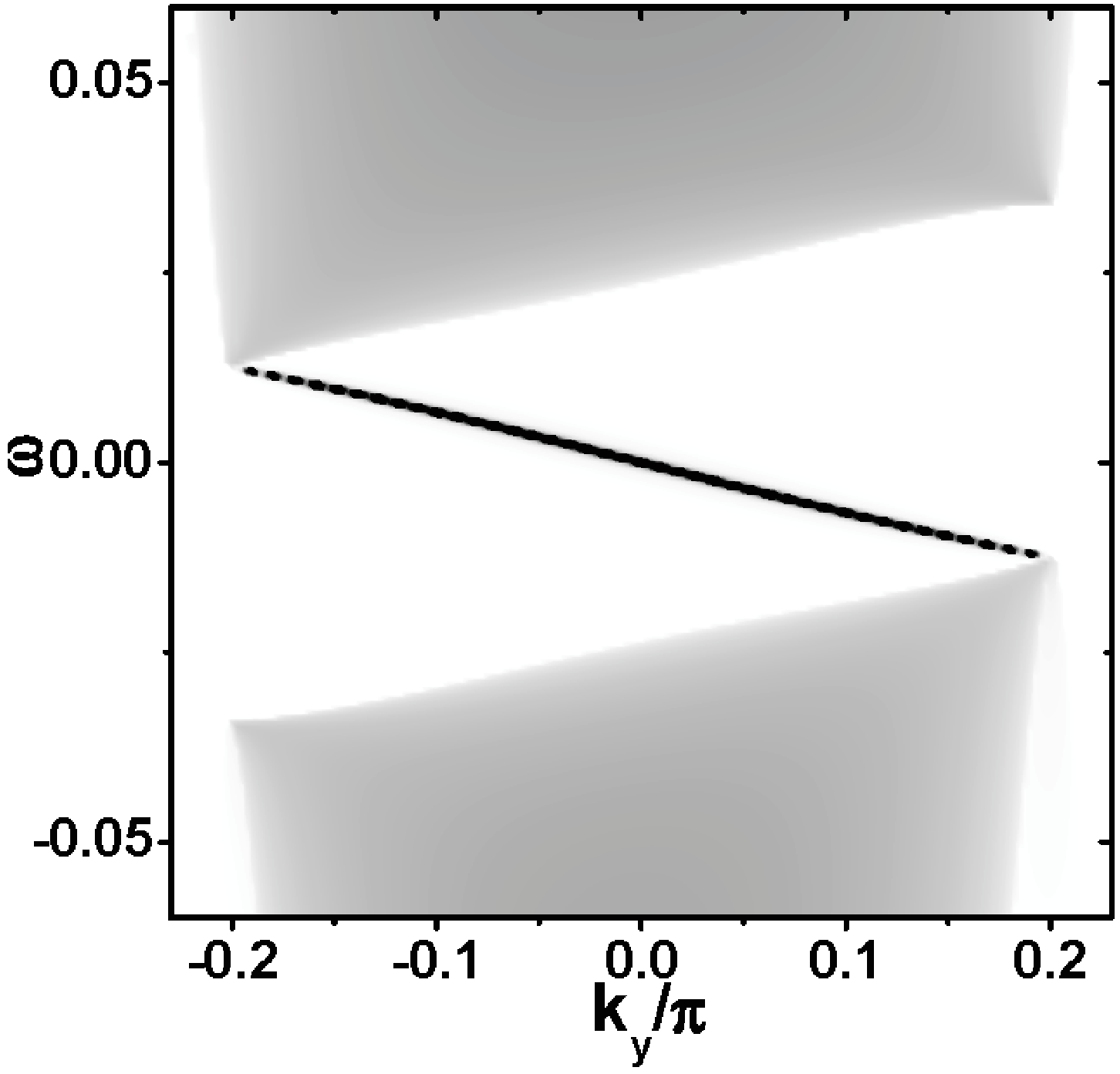}   \vspace{-0.10cm} \\
\hspace{-2.95cm} {\textbf{(e)}} \hspace{3.8cm}{\textbf{(f)}}\\
\hspace{0cm}\includegraphics[width=4.2cm,height=3.50cm]{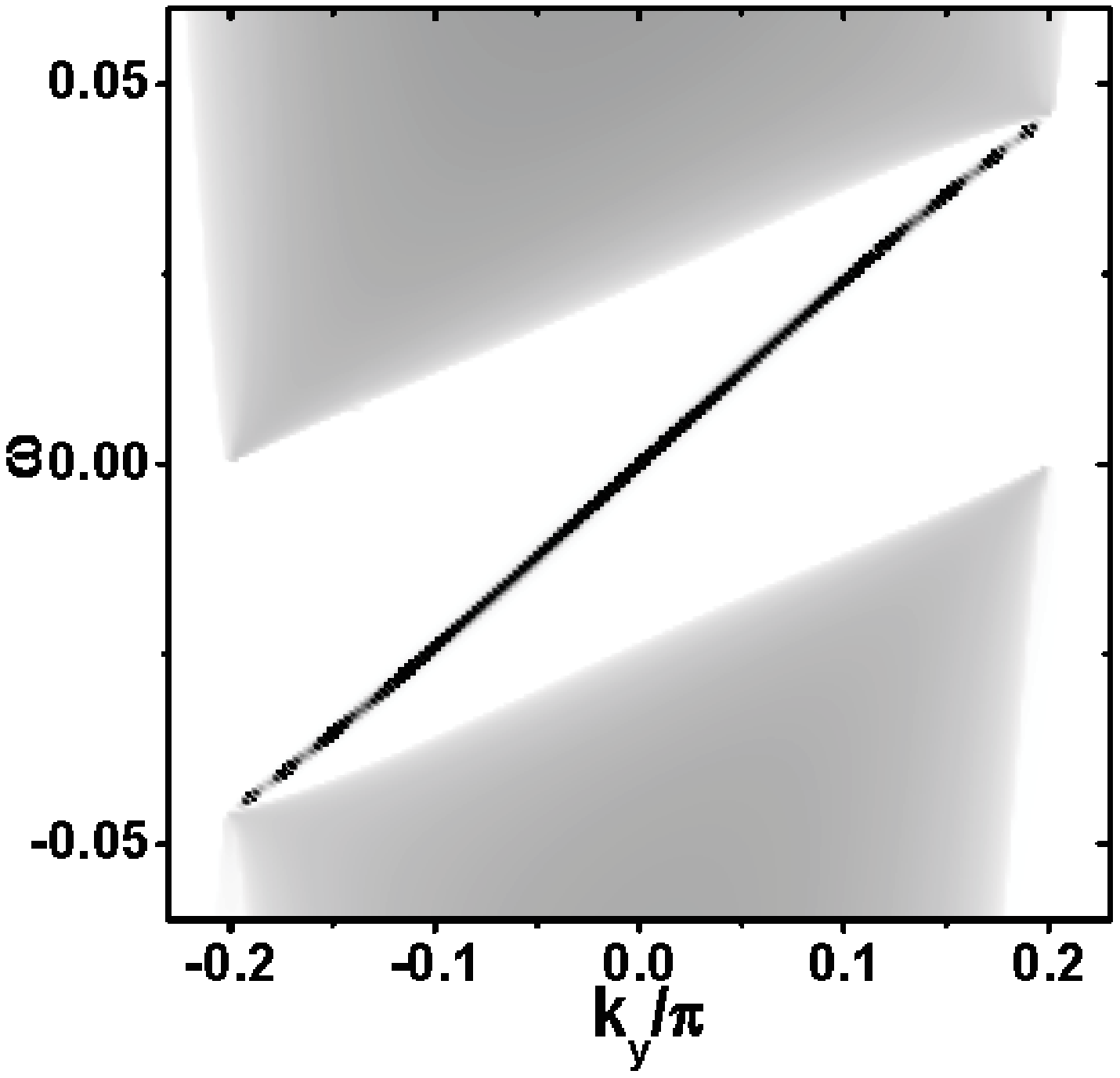}
\includegraphics[width=4.2cm,height=3.50cm]{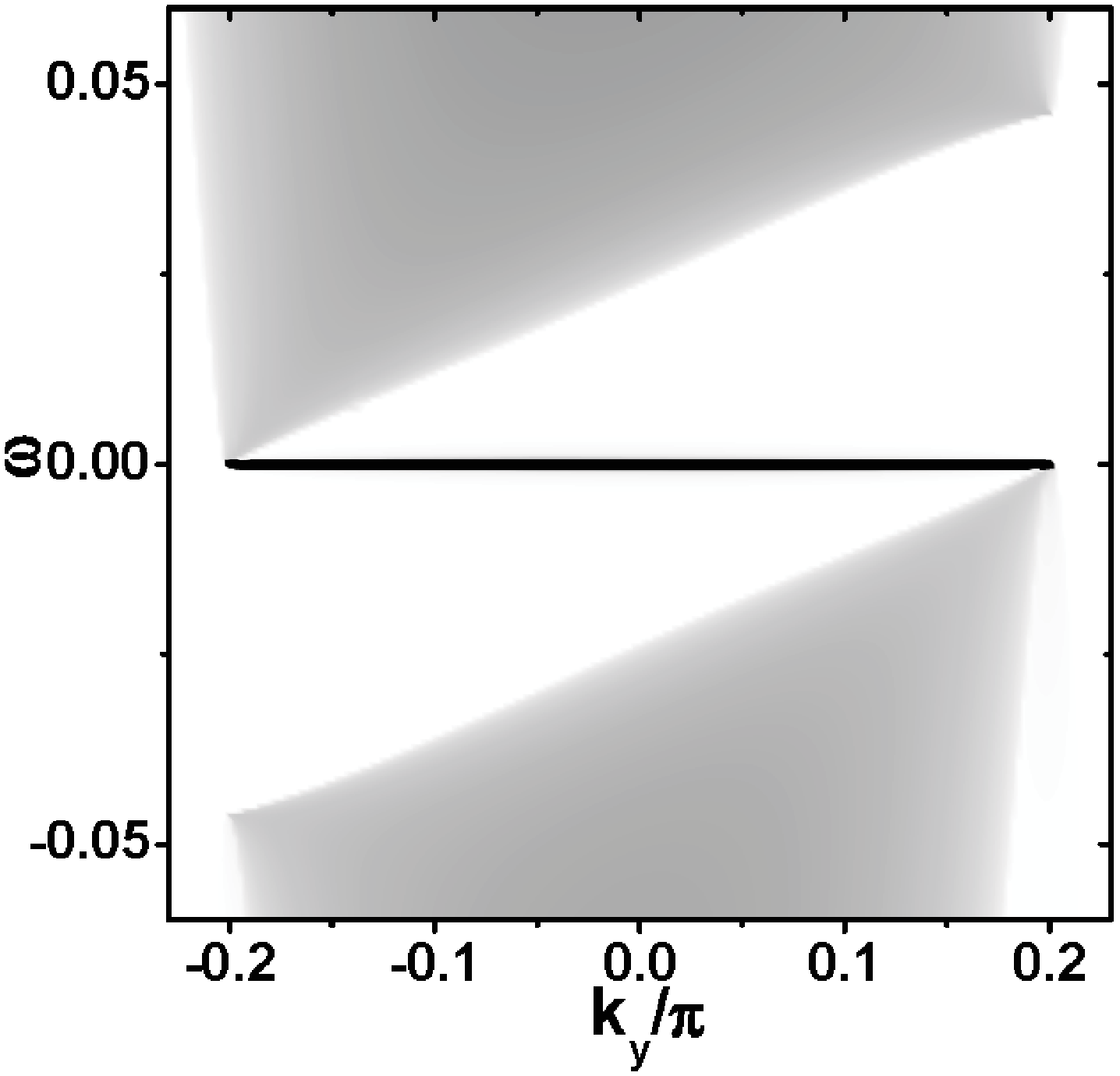}   \vspace{-0.10cm} \\
\hspace{-2.95cm}{\textbf{(g)}} \hspace{3.8cm}{\textbf{(h)}}\\
\hspace{0cm}\includegraphics[width=4.2cm,height=3.50cm]{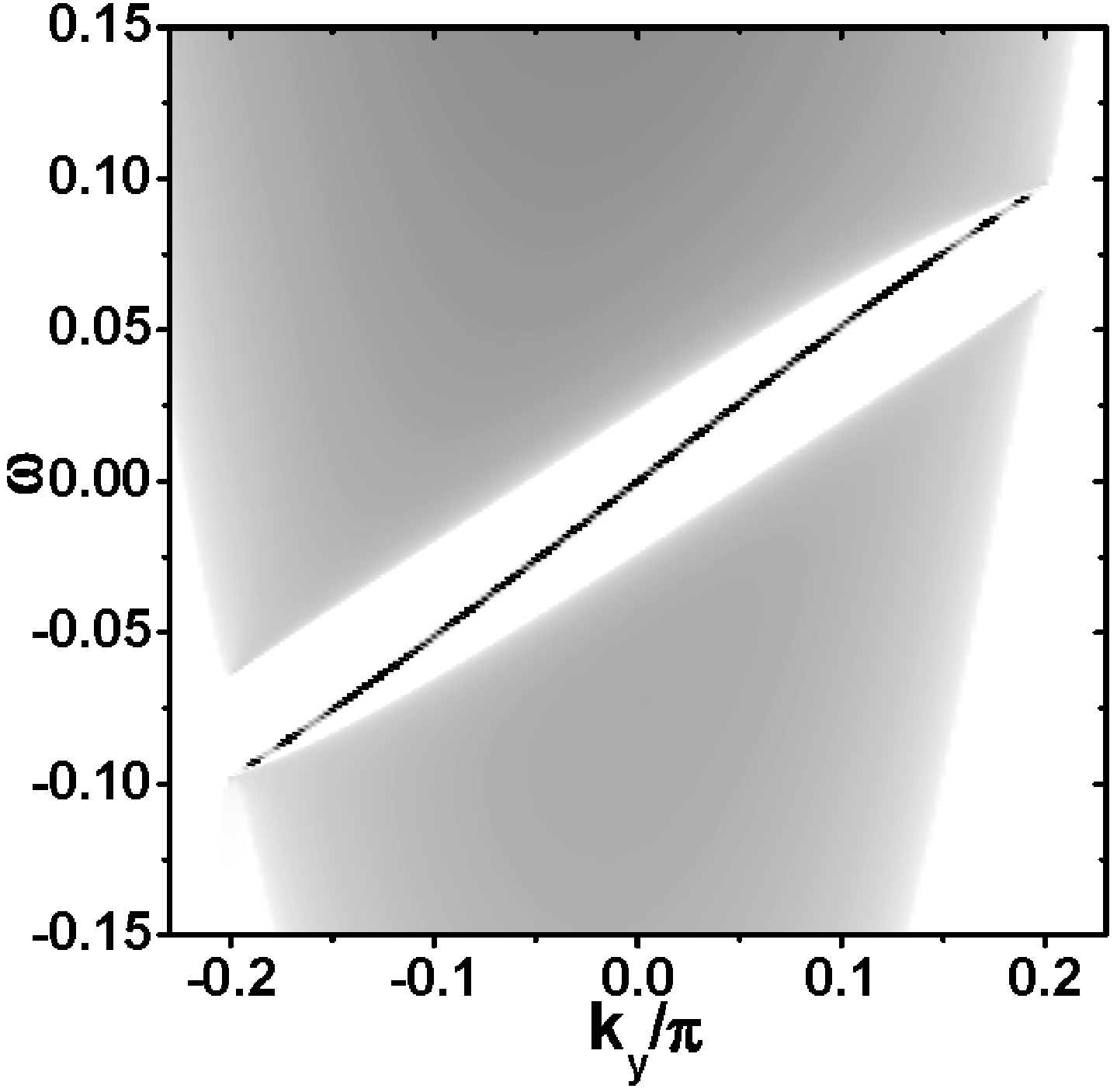}
\includegraphics[width=4.2cm,height=3.50cm]{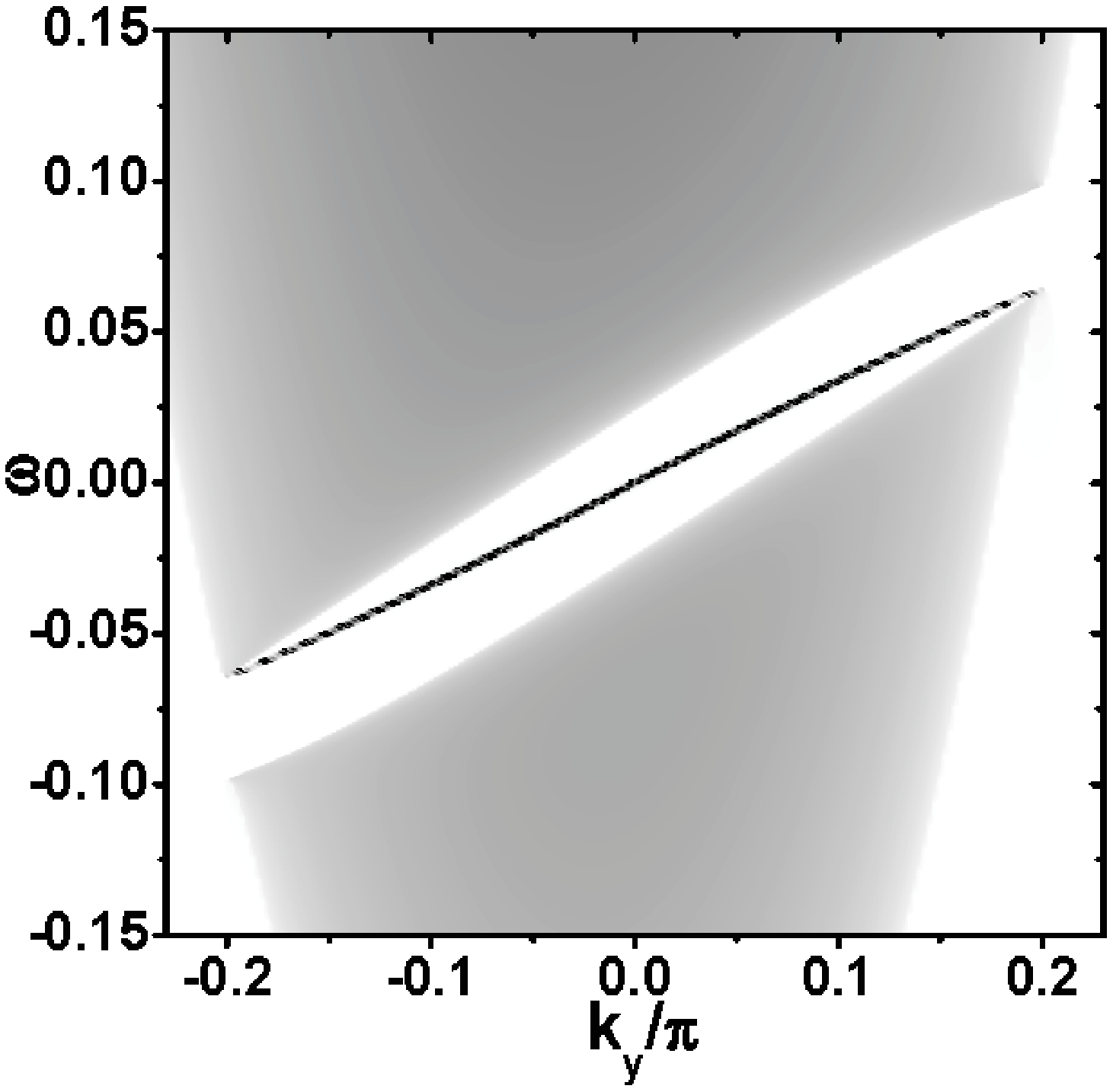}   \vspace{-0.10cm} \\
\hspace{-2.95cm}{\textbf{(i)}} \hspace{3.8cm}{\textbf{(j)}}\\
\hspace{0cm}\includegraphics[width=4.2cm,height=3.50cm]{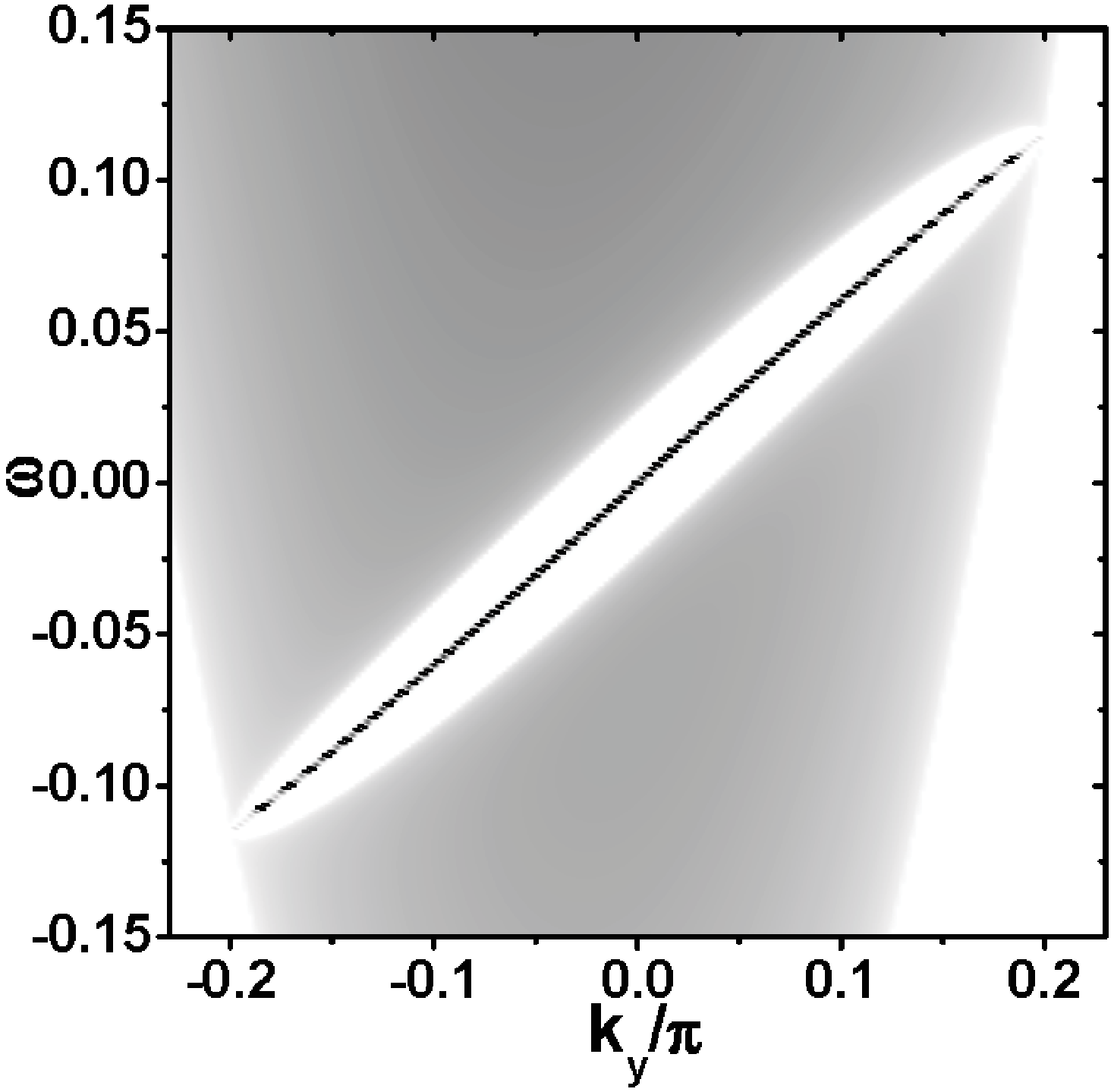}
\includegraphics[width=4.2cm,height=3.50cm]{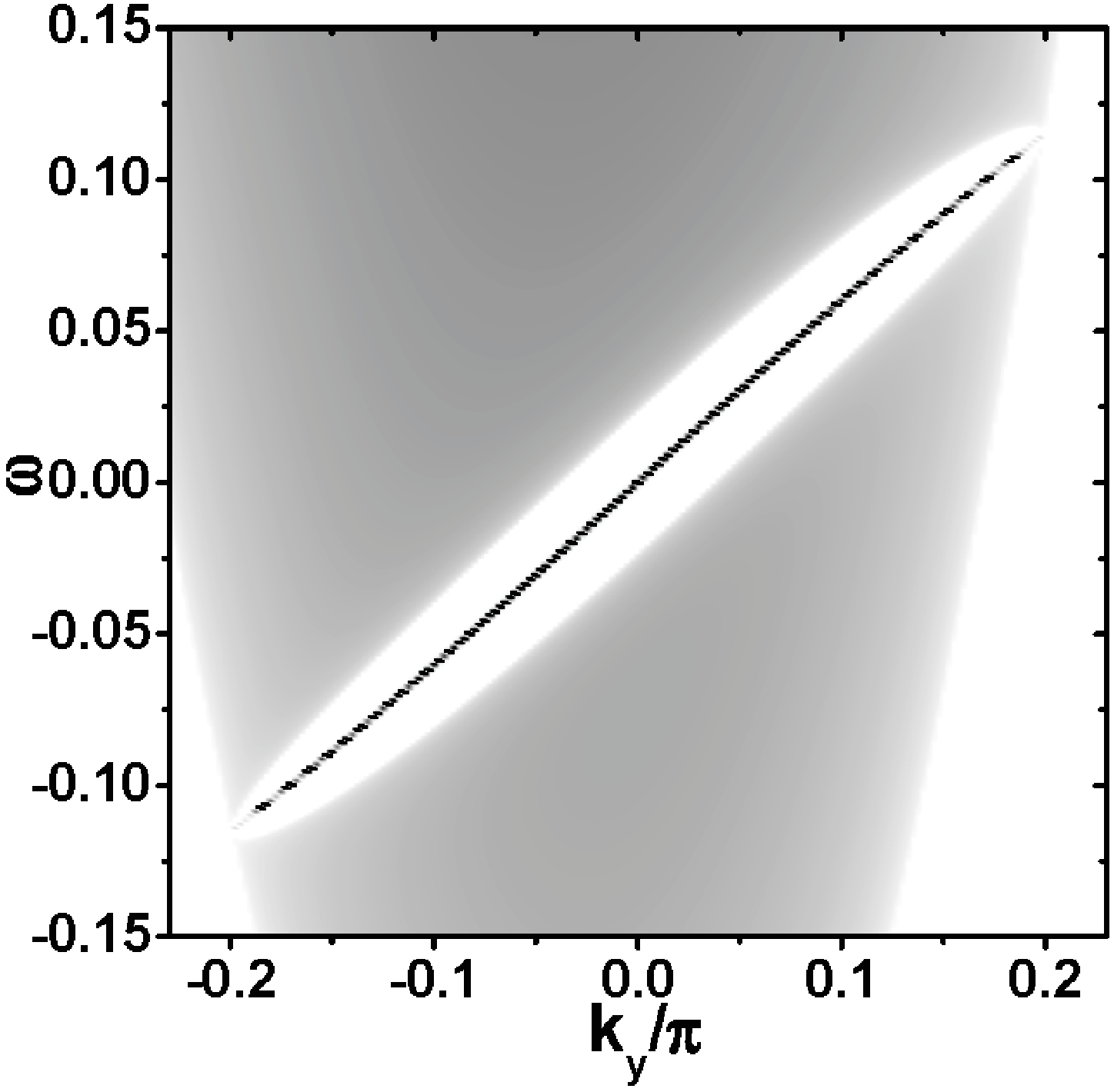}
\caption{Spectral functions for the left ($n_{x}=1$, in a, c, e, g, i) and right ($n_{x}=N_{x}$, in b, d, f, h, j) edges of a strip of the heterostructure. Five values of the angle are considered, including $\theta=0$ for (a) and (b), $\theta=0.03\pi$ for (c) and (d), $\theta=0.06415\pi$ for (e) and (f), $\theta=0.25\pi$ for (g) and (h), $\theta=0.5\pi$ for (i) and (j). The parameters are taken as $t>0$, $m=t$, $\lambda=0.2t$, $\Delta_{0}=0.2t$, and $\mu=-4.6t$. The energy $\omega$ is in unit of $t$. Darker color means larger spectral function.}
\end{figure}

As shown in Figure 5 are the spectral functions calculated in the above manner for the two edges ($L$ for $n_{x}=1$, $R$ for $n_{x}=N_{x}$). The model parameters ($t>0$, $m=t$, $\lambda=0.2t$, $\Delta_{0}=0.2t$, $\mu=-4.6t$) are the same as those used for Figures 2 and 3 of the main text. Results for five typical $\theta$ angles are displayed. In addition to the four angles considered for Figure 2 of the main text, we also include the results for $\theta=0$, for which the system is equivalent to the well-known $p+ip$ chiral superconductor, as was shown explicitly in Eq.(B8). From Figs. 5(a) to 5(d) for $\theta=0$ and $\theta=0.03\pi$, which correspond to the fully gapped phase with quantized TKNN number, the edge states on $n_{x}=1$ and $n_{x}=N_{x}$ have separately positive and negative velocities. As $\theta$ increases, the velocity of the edge states on $n_{x}=N_{x}$ decreases. When $\theta$ is increased to the critical angle $\theta_{c}$ ($\simeq0.06415\pi$ for the present parameters), the edge states on the $n_{x}=N_{x}$ edge becomes flat (Fig.5(f)). Increasing $\theta$ further, the system turns to the second phase (without a bulk gap but has a ubiquitous local gap) and the edge states on $n_{x}=1$ and $n_{x}=N_{x}$ become unidirectional and co-propagating (Figs. 5(g) and 5(h)). Finally, in the Weyl superconductivity phase (Figs. 5(i) and 5(j)), the two edge states have exactly the same dispersion and both connect the projections of the two bulk Weyl nodes on the edge BZ. Since the Weyl nodes are extended bulk states, this means that the two edge modes are connected together across the bulk of the strip through the two Weyl nodes.

\begin{figure}\label{fig6} \centering
\includegraphics[width=6.5cm,height=21cm]{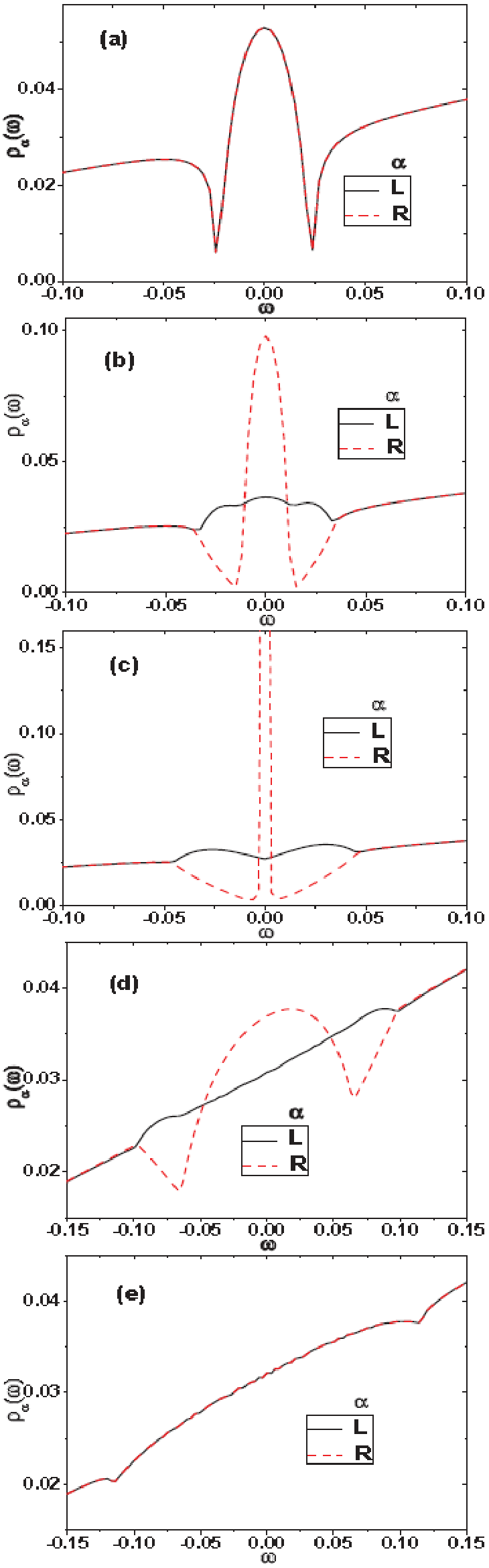}
\caption{Density of states on the two edges of a wide strip of the system, for five typical $\theta$ values. $\theta=0$ for (a), $\theta=0.03\pi$ for (b), $\theta=0.06415\pi$ for (c), $\theta=0.25\pi$ for (d), $\theta=0.5\pi$ for (e). The parameters are taken as $t>0$, $m=t$, $\lambda=0.2t$, $\Delta_{0}=0.2t$, and $\mu=-4.6t$. The energy $\omega$ is in unit of $t$. L and R represent the two edges with $n_{x}=1$ and $n_{x}=N_{x}$.}
\end{figure}

The spectral functions shown in Fig.5 are in principle observable by the angle-resolved photoemission spectroscopy (ARPES). Another promising method of probing the $\theta$-dependent edge states is the scanning tunneling spectroscopy (STS). For a clean sample with perfectly smooth and uniform edges, STS measures the DOS on the two edges, which are defined by Eqs.(C11) and (C12). The DOS of the edge states for five typical $\theta$ values which are the same as those used in Fig.5 are shown in Fig.6. For $\theta=0$, the broad peak in the DOS centering at $\omega=0$ can be understood as a combination of the constant (proportional to inverse of the velocity of the edge states) DOS from the 1D edge states with linear dispersion and the fact that the edge states penetrate more and more into the bulk of the strip as $(k_{y},\omega)$ deviates from $(0,0)$. As $\theta$ increases and approaches $\theta_{c}$, the edge states on $n_{x}=1$ ($n_{x}=N_{x}$) becomes increasing dispersive (flat), so the DOS on the right edge ($n_{x}=N_{x}$) gets enhanced as compared to the DOS on the left edge ($n_{x}=1$). The contrast between $\rho_{L}(\omega)$ and $\rho_{R}(\omega)$ attains its summit at $\theta=\theta_{c}$, when the edge states on the right edge becomes flat (Fig.6(c)). Then the difference between $\rho_{L}(\omega)$ and $\rho_{R}(\omega)$ decreases as $\theta$ increases further (Fig.6(d)). In the Weyl superconductivity phase (Fig.6(e)), the DOS on the edge layers are only slightly different from the DOS for the HM with RSOC in the normal phase. The two bulk Weyl nodes create two dips in the DOS of the edge states and the Fermi line connecting the two Weyl nodes gives a broad hump.

\section{proximity effect in a realistic setting}

To justify the simplified treatment of taking the proximity-induced pairing amplitude as a constant, namely $\Delta_{0}(\mathbf{k})=\Delta_{0}$, we consider a microscopic model for the interface between the HM and the $s$SC. The HM is still described by $\hat{H}_{0}$ defined in the main text. The $s$SC is also defined on a 2D square lattice and is assumed to match perfectly with the lattice of the HM. Defining the basis vector for the $s$SC as $\psi^{\dagger}_{\mathbf{k}}=[c^{\dagger}_{\mathbf{k}\uparrow},c^{\dagger}_{\mathbf{k}\downarrow}]$, the normal state is described by $\hat{H}'_{0}=\sum_{\mathbf{k}}\psi^{\dagger}_{\mathbf{k}}h'_{0}(\mathbf{k})\psi_{\mathbf{k}}$, where $h'_{0}(\mathbf{k})=\epsilon'_{\mathbf{k}}\sigma_{0}$. Up to nearest-neighbor (NN) hopping, $\epsilon'_{\mathbf{k}}=\epsilon_{0}-2t'(\cos k_{x}+\cos k_{y})-\mu$. $\epsilon_{0}$ measures the misalignment between the band centers of the HM and the $s$SC. The pairing term of the $s$SC is written as $\hat{H}'_{p}=\frac{1}{2}\sum_{\mathbf{k}}\psi^{\dagger}_{\mathbf{k}}\underline{\Delta}\psi^{\dagger}_{-\mathbf{k}}+\text{H.c.}$, where
$\underline{\Delta}=\Delta i\sigma_{2}$ with $\Delta$ a real constant number. Assuming perfect interface between the HM and the $s$SC, and assume the coupling occurs trough nearest-neighbor hopping along the direction perpendicular to the interface, we can model the coupling between the HM and the $s$SC with a tight-binding term as\cite{chung11,stanescu10}
\begin{equation}
\hat{H}_{mix}=\sum\limits_{\mathbf{k}}[\phi^{\dagger}_{\mathbf{k}}\gamma\sigma_{0}\psi_{\mathbf{k}}+\text{H.c.}],
\end{equation}
where $\gamma$ is a complex constant characterizing the strength of hybridization between the electronic wave unctions of the HM and the $s$SC. In the Nambu basis $\tilde{\psi}^{\dagger}_{\mathbf{k}}=[\psi^{\dagger}_{\mathbf{k}},\psi^{\text{T}}_{-\mathbf{k}}]$, the model for the $s$SC is written as
\begin{equation}
\hat{H}'=\frac{1}{2}\sum_{\mathbf{k}}\tilde{\psi}^{\dagger}_{\mathbf{k}}h'(\mathbf{k})\tilde{\psi}_{\mathbf{k}},
\end{equation}
where $h'(\mathbf{k})=\epsilon'_{\mathbf{k}}\tau_{3}\sigma_{0}-\Delta\tau_{2}\sigma_{2}$. The hybridization term is written as
\begin{equation}
\hat{H}_{mix}=\frac{1}{2}\sum\limits_{\mathbf{k}}[\varphi^{\dagger}_{\mathbf{k}}h_{t}\tilde{\psi}_{\mathbf{k}}+\text{H.c.}],
\end{equation}
where $h_{t}=(\text{Re}\gamma)\tau_{3}\sigma_{0}+i(\text{Im}\gamma)\tau_{0}\sigma_{0}$.

The effective pairing induced in the HM through proximity effect with the the $s$SC is contained in the following self-energy correction to the HM\cite{chung11,stanescu10}
\begin{eqnarray}
&&\Sigma(\mathbf{k},\omega)=h_{t}[\omega-h'(\mathbf{k})]^{-1}h^{\ast}_{t}  \\
&&=\frac{1}{\omega^{2}-\epsilon'^{2}_{\mathbf{k}}-\Delta^{2}}\begin{pmatrix} |\gamma|^{2}(\omega+\epsilon'_{\mathbf{k}})\sigma_{0} & i\gamma^{2}\Delta\sigma_{2}  \\
-i(\gamma^{\ast})^{2}\Delta\sigma_{2} & |\gamma|^{2}(\omega-\epsilon'_{\mathbf{k}})\sigma_{0}
\end{pmatrix}. \notag
\end{eqnarray}
Clearly, the proximity-induced pairing term is an even function of $\mathbf{k}$ and $\omega$, and is spin-singlet. The phase factor resulting from $\gamma^{2}$ is constant for all $\mathbf{k}$ and $\omega$ and thus is of no physical consequence. In the low-energy regime and for states close to the Fermi surface, the variation of the proximity-induced pairing amplitude is very small. As a result, we are qualitatively justified to work with an $s$-wave pairing term of constant pairing amplitude to model the proximity effect from the $s$SC, which gives Eq.(2) of the main text.

\end{appendix}


\end{document}